\documentclass[12pt]{article}

\usepackage{scicite}
\usepackage{graphicx}
\usepackage{times}
\usepackage{multibib}
\usepackage{lineno}

\newcommand{\grb}{GRB~130427A}
\newcommand{\h}{$^h$}
\newcommand{\m}{$^m$}
\newcommand{\s}{$^s$}
\newcommand{\D}{$^{\circ}$}
\newcommand{\am}{$'$}
\newcommand{\as}{$''$}
\newcommand{\sw}{{\it Swift}}
\newcommand{\frm}{{\it Fermi}}
\newcommand{\ra}{RA$_{J2000}$}
\newcommand{\dec}{Dec$_{J2000}$}
\newcommand{\cts}{cts s$^{-1}$}

\newcommand{\tg}{T$_{0, \rm GBM}$}

\newcommand{\tsta}{t$_{\rm{start}}$}
\newcommand{\tsto}{t$_{\rm{stop}}$}

\def\gsim{ \lower .75ex \hbox{$\sim$} \llap{\raise .27ex \hbox{$>$}} }
\def\lsim{ \lower .75ex\hbox{$\sim$} \llap{\raise .27ex \hbox{$<$}} }

\newenvironment{sciabstract}{%
\begin{quote} \bf}
{\end{quote}}

\newcounter{lastnote}
\newenvironment{scilastnote}{%
\setcounter{lastnote}{\value{enumiv}}%
\addtocounter{lastnote}{+1}%
\begin{list}%
{\hspace{1mm}}
{\setlength{\leftmargin}{.22in}}
{\setlength{\labelsep}{.5em}}}
{\end{list}}

\title{\grb: a Nearby Ordinary Monster}

\author
{A.~Maselli$^{1\ast}$, A.~Melandri$^{2}$, L.~Nava$^{2,3}$, C.G.~Mundell$^{4}$, N.~Kawai$^{5,6}$,\\ 
 S.~Campana$^{2}$, S.~Covino$^{2}$, J.R.~Cummings$^{7}$, G.~Cusumano$^{1}$, P.A.~Evans$^{8}$,\\
 G.~Ghirlanda$^{2}$, G.~Ghisellini$^{2}$, C.~Guidorzi$^{9}$, S.~Kobayashi$^{4}$, P.~Kuin$^{10}$,\\ 
 V.~La~Parola$^{1}$, V.~Mangano$^{1,11}$, S.~Oates$^{10}$, T.~Sakamoto$^{12}$, M.~Serino$^{6}$,\\ 
 F.~Virgili$^{4}$, B.-B.~Zhang$^{11}$, S.~Barthelmy$^{13}$, A.~Beardmore$^{8}$, M.G.~Bernardini$^{2}$,\\ 
 D.~Bersier$^{4}$, D.~Burrows$^{11}$, G.~Calderone$^{2,14}$, M.~Capalbi$^{1}$, J.Chiang$^{15}$,\\ 
 P.~D'Avanzo$^{2}$, V.~D'Elia$^{16,17}$, M.~De~Pasquale$^{10}$, D.~Fugazza$^{2}$, N.~Gehrels$^{13}$,\\ 
 A.~Gomboc$^{18,19}$, R.~Harrison$^{4}$, H.~Hanayama$^{20}$, J.~Japelj$^{18}$, J.~Kennea$^{11}$,\\ 
 D.~Kopac$^{18}$, C.~Kouveliotou$^{21}$, D.~Kuroda$^{22}$, A.~Levan$^{23}$, D.~Malesani$^{24}$,\\ 
 F.~Marshall$^{13}$, J.~Nousek$^{11}$, P.~O'Brien$^{8}$, J.P.~Osborne$^{8}$, C.~Pagani$^{8}$,\\ 
 K.L.~Page$^{8}$, M.~Page$^{10}$, M.~Perri$^{16,17}$, T.~Pritchard$^{11}$, P.~Romano$^{1}$, Y.~Saito$^{5}$,\\ 
 B.~Sbarufatti$^{2,11}$, R.~Salvaterra$^{25}$, I.~Steele$^{4}$, N.~Tanvir$^{8}$, G.~Vianello$^{15}$,\\ 
 B.~Wiegand$^{13}$, K.~Wiersema$^{8}$, Y.~Yatsu$^{5}$, T.~Yoshii$^{5}$, \& G.~Tagliaferri$^{2}$\\
\\
\normalsize{$^\ast$To whom correspondence should be addressed; E-mail: maselli@ifc.inaf.it}
}

\date{}

\begin{document} 

\baselineskip24pt

\maketitle 

\noindent{\footnotesize{$^{1}$INAF-IASF Palermo, Via Ugo La Malfa 153 I-90146 Palermo, Italy}; \footnotesize{$^{2}$INAF-Osservatorio Astronomico di Brera, via E. Bianchi 46, I-23807 Merate, Italy}; \footnotesize{$^{3}$APC, Universit\'e Paris Diderot, CNRS/IN2P3, CEA/Irfu, Observatoire de Paris, Sorbonne Paris Cit\'e, France}; \footnotesize{$^{4}$Astrophysics Research Institute, Liverpool John Moores University, Liverpool Science Park, 146 Brownlow Hill, Liverpool L3 5RF, UK}; \footnotesize{$^{5}$Department of Physics, Tokyo Institute of Technology, 2-12-1 Ookayama, Meguro-ku, Tokyo 152-8551, Japan}; \footnotesize{$^{6}$Coordinated Space Observation and Experiment Research Group, RIKEN, 2-1 Hirosawa, Wako, Saitama 351-0198, Japan}; \footnotesize{$^{7}$UMBC/CRESST/NASA Goddard Space Flight Center, Code 661, Greenbelt, MD 20771, USA}; \footnotesize{$^{8}$Department of Physics and Astronomy, University of Leicester, Leicester, LE1 7RH, UK};  \footnotesize{$^{9}$Department of Physics, University of Ferrara, via Saragat 1, I-44122, Ferrara, Italy}; \footnotesize{$^{10}$Mullard Space Science Laboratory, University College London, Holmbury St. Mary, Dorking, Surrey RH5 6NT, UK}; \footnotesize{$^{11}$Department of Astronomy \& Astrophysics, The Pennsylvania State University, 525 Davey Lab, University Park, PA 16802, USA}; \footnotesize{$^{12}$Department of Physics and Mathematics, Aoyama Gakuin University, 5-10-1 Fuchinobe, Chuo-ku, Sagamihara, Kanagawa 252-5258, Japan}; \footnotesize{$^{13}$NASA Goddard Space Flight Center, Greenbelt, MD 20771, USA}; \footnotesize{$^{14}$Dipartimento di Fisica ``G. Occhialini'', Universit\`a di Milano-Bicocca, Piazza della Scienza 3, I-20126 Milano, Italy}; \footnotesize{$^{15}$W. W. Hansen Experimental Physics Laboratory, Kavli Institute for Particle Astrophysics and Cosmology, Department of Physics, and SLAC National Accelerator Laboratory, Stanford University, Stanford, CA 94305, USA; \footnotesize{$^{16}$INAF/Rome Astronomical Observatory, via Frascati 33, 00040 Monteporzio Catone (Roma), Italy}; \footnotesize{$^{17}$ASI Science Data Centre, Via Galileo Galilei, 00044 Frascati (Roma), Italy}; \footnotesize{$^{18}$Faculty of Mathematics and Physics, University of Ljubljana, Jadranska 19 1000, Ljubljana, Slovenia}; \footnotesize{$^{19}$Centre of Excellence Space-si, A\v{s}ker\v{c}eva cesta 12, 1000 Ljubljana, Slovenia}; \footnotesize{$^{20}$Ishigakijima Astronomical Observatory, National Astronomical Observatory of Japan, 1024-1 Arakawa, Ishigaki, Okinawa 907-0024, Japan}; \footnotesize{$^{21}$Space Science Office, VP62, NASA/Marshall Space Flight Center, Huntsville, AL 35812, USA}; \footnotesize{$^{22}$Okayama Astrophysical Observatory, National Astronomical Observatory of Japan, 3037-5 Honjo, Kamogata, Asaguchi, Okayama 719-0232}; \footnotesize{$^{23}$Department of Physics, University of Warwick, Coventry CV4 7AL, UK}; \footnotesize{$^{24}$Dark Cosmology Centre, Niels Bohr Institute, University of Copenhagen, Juliane Maries Vej 30, 2100 Copenhagen, Denmark}; \footnotesize{$^{25}$INAF-IASF Milano, via E. Bassini 15, I-20133 Milano, Italy}}}

\newpage

\begin{sciabstract}

Long-duration Gamma-Ray Bursts (GRBs) are an extremely rare outcome of
the collapse of massive stars, and are typically found in the distant
Universe.
Because of its intrinsic luminosity ($L\sim 3 \times 10^{53}$ erg
s$^{-1}$) and its relative proximity ($z=0.34$), \grb\ was a unique
event that reached the highest fluence observed in the $\gamma$-ray
band.
Here we present a comprehensive multiwavelength view of \grb\ with
\sw, the 2-m Liverpool and Faulkes telescopes and by other
ground-based facilities, highlighting the evolution of the burst emission
from the prompt to the afterglow phase.
The properties of \grb\ are similar to those of the most luminous,
high-redshift GRBs, suggesting that a common central engine is
responsible for producing GRBs in both the contemporary and the early
Universe and over the full range of GRB isotropic energies.
\end{sciabstract}

\section*{}
\label{sect:main}

\grb\ was the brightest burst detected by
\sw\ \cite{2013GCN..14448...1M} as well as by several $\gamma$--ray
detectors onboard other space missions.
It was also the brightest and longest burst detected above 100~MeV,
with the most energetic photon detected at 95~GeV \cite{LAT}.
It was detected by \frm-GBM \cite{2009ApJ...702..791M} at \tg\ =
07:47:06.42~UT on April~27~2013.
Hereafter this time will be our reference time T$_0$.
The Burst Alert Telescope (BAT, \cite{2005SSRv..120..143B}) onboard
\sw\ triggered on \grb\ at t~=~51.1~s, when \sw\ completed a
pre-planned slew.
The \sw\ slew to the source started at t~=~148~s and ended at
t~=~192~s.
The \sw\ UltraViolet Optical Telescope (UVOT,
\cite{2005SSRv..120...95R}) began observations at t~=~181~s while
observations by the \sw\ X--ray Telescope (XRT,
\cite{2005SSRv..120..165B}) started at t~=~195~s (see \cite{som} for
more details).
The structure of the $\gamma$-ray light curve revealed by the \sw-BAT
in the 15--350~keV band (Fig.~1) can be divided in three main
episodes: an initial peak, beginning at t~=~0.1~s and peaking at
t~=~0.5~s; a second large peak showing a complex structure with a
duration of $\sim$ 20~s and a third, much weaker episode, starting at
t~$\sim$120~s showing a fast rise/exponential decay behavior.
The overall duration of the prompt emission was T$_{90}$ (15$-$150
keV) = 276~$\pm$~5~s (i.e. the time containing $90\%$ of the fluence)
calculated over the first 1830~s of BAT observation from \tg.
During the early phases of the $\gamma$-ray emission strong spectral
variability is observed (Fig.~1).
A marked spectral hardening is observed during the prompt main event.
With a total fluence $F = (4.985\pm0.002)\times 10^{-4}$ erg~cm$^{-2}$
in the 15--150~keV band, \grb\ reached the highest fluence observed
for a GRB by \sw.
The 0.02--10~MeV fluence measured by Konus-Wind
\cite{1995SSRv...71..265A} for the main emission episode (0~--~18.7~s)
is $(2.68~\pm~0.01) \times 10^{-3}$ erg cm$^{-2}$, with a spectrum
peaking at $E_{\rm peak} = 1028~\pm~8$~keV, while the fluence of the
emission episode at (120~--~250~s) is $\sim 9 \times 10^{-5}$ erg
cm$^{-2}$, with a spectrum peaking at $\sim$240~keV
\cite{2013GCN..14487...1G}.

This event was extremely bright also in the optical and it was
immediately detected by various robotic telescopes: in particular, the
Raptor robotic telescope detected a bright optical counterpart already
at t~=~0.5~s \cite{RAPTOR}.
Optical spectroscopy of the afterglow determined the redshift to be
$z$~=~0.34 \cite{2013GCN..14455...1L}; an UVOT UV grism spectrum
\cite{som} was also acquired.
At this distance the rest frame 1~keV--10~MeV isotropic energy is
$E_{\rm iso} = 8.1 \times 10^{53}$~erg and the peak luminosity is
$L_{\rm iso} = 2.7 \times 10^{53}$~erg~s$^{-1}$.
According to the luminosity function of Salvaterra et
al. \cite{2012ApJ...749...68S} we expect one event like \grb\ every
$>60$ years.
In the nearby Universe (i.e. $z~\lsim~0.4$, corresponding to an age of
$\sim 10$ Gyr) only a handful of long GRBs have been detected.
These GRBs are usually characterized by a low overall isotropic energy
(E$_{\rm iso} \le 10^{52}$ erg) and are associated with supernovae (SNe) Ib/c,
characterized by broad spectral lines indicating high expansion velocities,
called hypernovae \cite{2006ARA&A..44..507W}.
\grb\ is instead a powerful GRB such as the ones typically detected at
much higher redshifts ($z~>~1$, with a mean $z \sim 2$ corresponding
to an age of $\sim 3$ Gyr).
The detection of a nearby and extremely powerful GRB gives us the
opportunity to test, on the one hand, if this GRB has the same
properties of the cosmological GRBs and, on the other hand, if also
such bright GRBs are associated with SNe. 
Up to now, SNe have been associated only to under (or mildly)
energetic GRBs in the local Universe.
Since a supernova associated with this burst, SN~2013cq, has been
detected \cite{2013ApJ...776...98X}, we are now sure that SNe are also
associated with very powerful GRBs, not only to low power bursts (see
Fig.~S6; see also Fig.~1 of \cite{2013ApJ...776...98X}).
Naive energetic arguments might suggest that in powerful GRBs there is
not enough power left for a strong SN: \grb\ definitively proves that
this is not the case.

The overall behavior of the X--ray afterglow light curve has been
characterized with the main contribution of the XRT onboard \sw\ and
two additional relevant detections from the MAXI experiment
\cite{2009PASJ...61..999M} in the gap between the first and the second
\sw-XRT observations (Fig.~2).
The early light curve, starting from t~=~260~s, is characterized by an
initially steep decay with a slope $\alpha_{0, \rm x}$~=~3.32~$\pm$~0.17 
consistent with high-latitude emission 
\cite{2005Natur.436..985T,2006ApJ...642..389N}, a break at 
t$_{1,\rm x}$~=~424~$\pm$~8~s followed by a flatter decay with index 
$\alpha_{1, \rm x}$~=~1.28~$\pm$~0.01.
A further break at t$_{2, \rm x}$~=~48~$\pm$~22~ks is statistically
needed ($3.8 \sigma$) to account for a further steepening to
$\alpha_{2, \rm x}$~=~1.35~$\pm$~0.02 (all errors are derived for
$\Delta\chi^2=2.7$).

Figure~2 also shows the light curves in the optical and UV derived
from the UVOT as well as from ground-based telescopes (Liverpool
telescope, Faulkes Telescope North, and MITSuME Telescopes).
All optical light curves are well fitted by a broken power law with
$\alpha_1$~=~0.96~$\pm$~0.01, t$_{\rm
  {break}}$~=~37.4~$^{+4.7}_{-4.0}$~ks and
$\alpha_2$~=~1.36~$^{+0.01}_{-0.02}$.
Fitting the X--ray light curve together with the optical ones, we find
the same parameters from 26.6~ks onward, but to fit the early part of
the X--ray light curve we need another power--law segment with a slope of
1.29~$^{+0.02}_{-0.01}$ and a break at 26.6$^{+4.5}_{-6.6}$~ks (Fig.~2).
Therefore, from 26.6~ks onward a common description of all the optical, UV
and X--ray behavior is possible, while at earlier times an extra X--ray
component is required.

We interpret the early X--ray light curve (up to 26.6~ks) as the
superposition of a standard afterglow (i.e. forward shock emission)
and either the prolonged activity of the central engine or/and the
contribution from the reverse shock emission
(e.g. \cite{2006RPPh...69.2259M,2009MNRAS.393..253G,2011MNRAS.414.3537P}).
After 26.6~ks the optical and X--ray light curves share the same
behavior and decay slopes (Fig.~2), including a break at
$t_{\rm break} \sim 37$~ks.
This achromatic break is suggestive of a jet break, although the post-break
decay ($\alpha_2=1.36$) is shallower than predicted in the simplest theory
(an increase in decay slope $>1$ would be expected; see
\cite{1999ApJ...519L..17S}).
This could be due to additional components contributing to the flux,
to a time dependence of the microphysical parameters governing the
fraction of shock energy going to electrons ($\epsilon_{\rm e}$) and
magnetic field ($\epsilon_{\rm B}$), or to the fact that we observe a
canonical jet not exactly on axis, but still within the jet opening
angle \cite{2010ApJ...722..235V,2012ApJ...751..155V}.

Because the optical and the X--ray emission belong to the same spectral
power--law segment it is possible to constrain the characteristic
frequencies of the afterglow spectra, in turn constraining the
microphysical parameters of the relativistic shock.
Additional information comes from the high-energy $\gamma$-ray
emission \cite{LAT}.
The $\gamma$--ray flux above 100~MeV peaks at $\sim 20$~s.
If this emission is due to afterglow radiation, the peak time implies
a bulk Lorentz factor $\Gamma_0~\sim~500$ \cite{som,LAT}.
Furthermore, the presence of the GeV peak suggests a homogeneous
circumburst density profile \cite{2013MNRAS.433.2107N}.
Guided by these constraints in our choice of parameters, we used the
{\tt BOXFIT} code developed by van Eerten et
al. \cite{2012ApJ...749...44V} to model the afterglow.
Rather than trying to perform a formal fit to the data we check if
this burst, with an unprecedented data coverage and richness, can be
interpreted in the framework of the standard model for the afterglow
emission, or else if it forces us to abandon the standard framework.
We give more weight to the optical and higher energy fluxes, since
they carry most of the afterglow luminosity, orders of magnitude
greater than the radio flux.

Neither reverse shock nor inverse Compton (IC) emission are included
in the model, but this does not affect our conclusions which primarily
concern the late time synchrotron afterglow (see \cite{som} for
further details).
The synchrotron flux predicted by the model reproduces the optical
emission and the X--ray light curve after $\sim 10$~ks reasonably
well, while the early X--ray flux is likely due to an additional
component (Fig.~3).
Our model underestimates the GeV emission but this, given the large
$\epsilon_{\rm e}/\epsilon_{\rm B}$ ratio, can be due to synchrotron
Self--Compton emission, as envisaged by
\cite{2013ApJ...773L..20L,2013ApJ...771L..13T,LAT}.
The model can also roughly reproduce the radio emission (further
details in \cite{som}).

Consistent with the light curve analysis, the synchrotron flux
predicted by the model reproduces reasonably well the optical and
X--ray parts of the spectral energy distribution (SED) of the
afterglow (Fig.~4), but it underestimates the GeV emission.
Although the model does not entirely reproduce the complexity shown by
the data, it does capture the main features of the emission properties
in the pure afterglow phase.

Overall, the properties of \grb\ are similar to those of the powerful
GRBs typically seen at $z \sim 1-2$ (see \cite{som} for a comparison).
This is the most powerful GRB at $z<0.9$.
It obeys the spectral energy correlations such as the $E_{\rm peak} -
E_{\rm iso}$ \cite{2002A&A...390...81A} correlation and the $E_{\rm
  peak} - L_{\rm peak}$ \cite{2004ApJ...609..935Y} correlation.
Interpreting the break observed at $\sim 37$~ks as a jet break makes
\grb\ consistent with the collimation corrected energy--peak energy
correlation \cite{som,2004ApJ...616..331G}.
\grb\ is also associated with a supernova, extending the GRB-SN
connection also to such powerful and high--$z$ bursts.
\grb\ stands as a unique example indicating that a common engine is
powering these huge explosions at all powers, and from the
nearby to the very far, early Universe.

\nocite{2013ApJ...776..119L}
\nocite{2013GCN..14473...1V}
\nocite{2013GCN..14494...1P}
\nocite{2006A&A...456..917R}
\nocite{2000SPIE.4140...76R}
\nocite{2004SPIE.5165..262R}
\nocite{2009MNRAS.395..490O}
\nocite{2013MNRAS.436.1684P}
\nocite{2010MNRAS.406.1687B}
\nocite{2011AIPC.1358..373B}
\nocite{2008MNRAS.383..627P}
\nocite{2011PASJ...63S.623M}
\nocite{2011PASJ...63S.635S}
\nocite{2007ApJ...663..407B}
\nocite{2009MNRAS.397.1177E}
\nocite{2013MNRAS.428..729M}
\nocite{2006PASP..118..288G}
\nocite{2012IAUS..279..387S}
\nocite{1997ApJ...489L..37S}
\nocite{2012MNRAS.420..483G}
\nocite{2003ApJ...597..455K}
\nocite{2014ApJ...781...37P}
\nocite{2001ApJ...560L..49P}
\nocite{2010MNRAS.409..226K}
\nocite{2010MNRAS.403..926G}
\nocite{2004MNRAS.354...86R}
\nocite{2010ApJ...718L..63P}
\nocite{2011MNRAS.410.2016V}
\nocite{2006NCimB.121.1099P}
\nocite{2012MNRAS.421.1256N}
\nocite{1998ApJ...500..525S}
\nocite{1996AJ....111.1748F}

\newpage

\bibliography{scibib}

\bibliographystyle{Science}

\begin{scilastnote}
\item {\bf Acknowledgements}: This work has been supported by ASI
  grant I/004/11/0 and by PRIN-MIUR grant 2009ERC3HT. Development of
  the Boxfit code \cite{2012ApJ...749...44V} was supported in part by
  NASA through grant NNX10AF62G issued through the Astrophysics Theory
  Program and by the NSF through grant AST-1009863. This research was
  partially supported by the Ministry of Education, Culture, Sports,
  Science and Technology of Japan (MEXT), Grant-in-Aid No. 14GS0211,
  19047001, 19047003, and 24740186. The Liverpool Telescope is
  operated by Liverpool John Moores University at the Observatorio del
  Roque de los Muchachos of the Instituto de Astrof\'{i}sica de
  Canarias. The Faulkes Telescopes, now owned by Las Cumbres
  Observatory, are operated with support from the Dill Faulkes
  Educational Trust. \sw\ support at the University of Leicester and
  the Mullard Space Science Laboratory is funded by the UK Space
  Agency. C.G. Mundell acknowledges financial support from the Royal
  Society, the Wolfson Foundation and the Science and Technology
  Facilities Council. Andreja Gomboc acknowledges funding from the
  Slovenian Research Agency and from the Centre of Excellence for
  Space Sciences and Technologies SPACE-SI, an operation partly
  financed by the European Union, European Regional Development Fund
  and Republic of Slovenia. DARK is funded by the DNRF.
\end{scilastnote}

\newpage 

\begin{figure}[hbtp]
\begin{center}
\hspace{-1.5cm}
\includegraphics[width=16cm,]{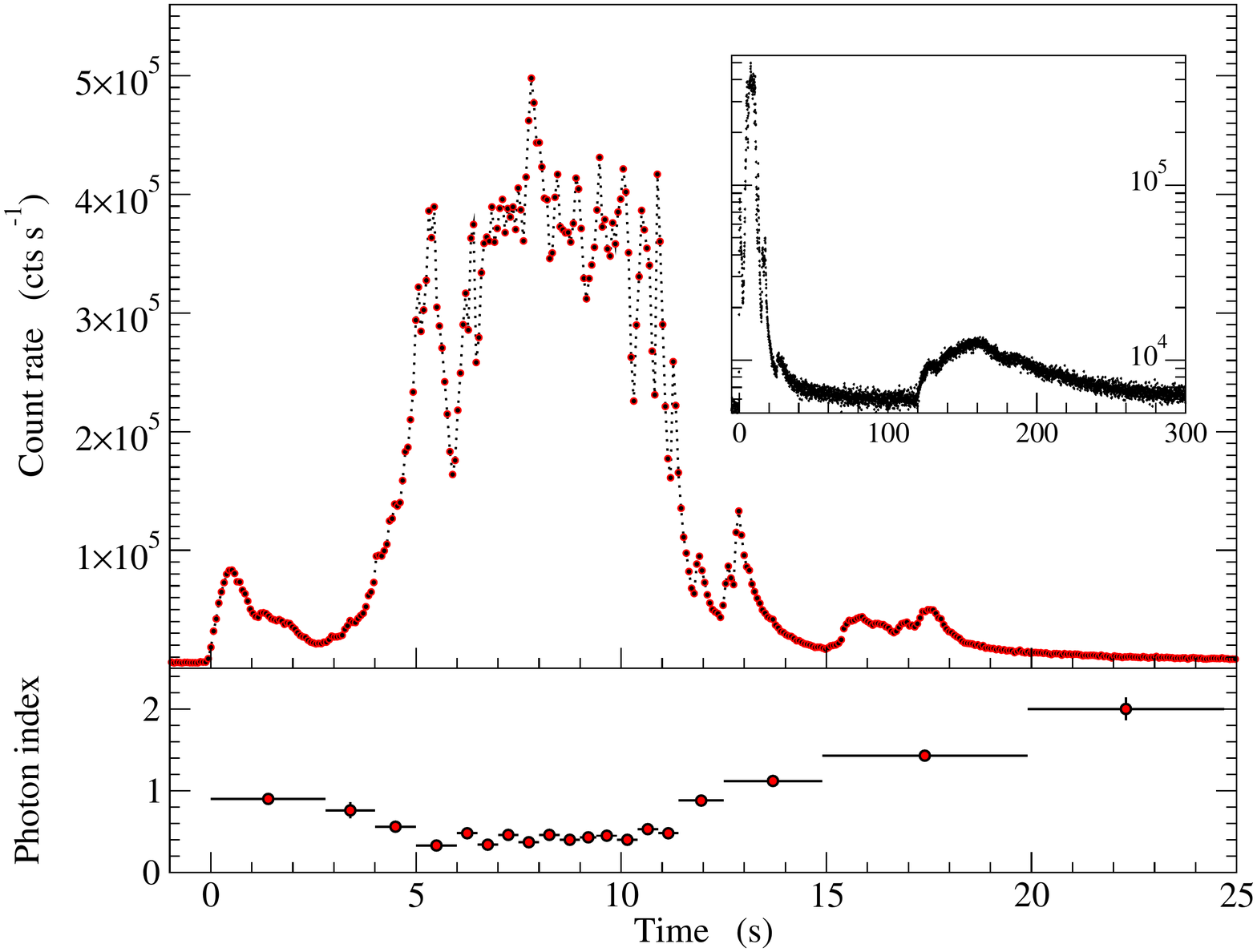}
\end{center}
\end{figure}
\vskip -1 cm
\noindent{\footnotesize{\bf Figure~1.} Details of the \sw\ BAT light
  curve in the 15--350 keV band. Top panel: the BAT light curve with a
  binning time of 64~ms. Inset: the BAT light curve up to 300~s,
  plotted on a log intensity scale, showing a fast rise/exponential
  decay feature starting at t~$\sim$~120~s. Bottom panel: the photon
  index values of a power-law model fit to the BAT spectrum in the
  15--150 keV energy range.}

\newpage 

\begin{figure}[hbtp]
\begin{center}
\vskip -1.5 cm
\includegraphics[width=16.5cm]{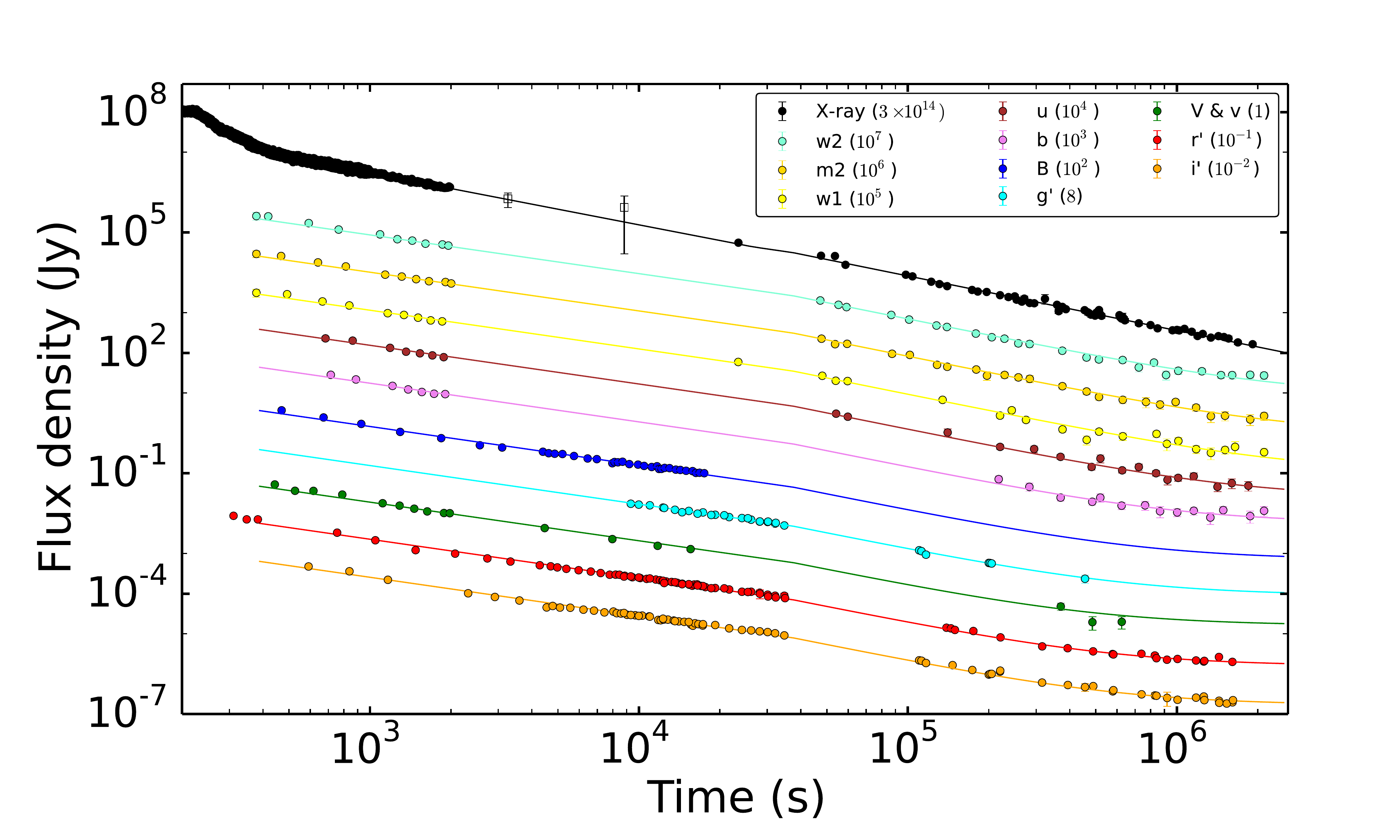}
\end{center}
\end{figure}
\vskip -1 cm
\noindent{\footnotesize{\bf Figure~2}: Light curves for \grb\ in
  different wavebands.  \sw\ UV and visible filters ($w2$, $m2$, $w1$,
  $u$, $b$, $v$); $B$, $V$, $r'$, and $i'$ filters correspond to
  Faulkes Telescope North; $r'$, and $i'$ to the Liverpool Telescope;
  $g'$, $r'$, and $i'$ to the MITSuME telescopes (see {\it 7} for
  further details). The scaling factors for the flux density in
  different filters is shown in the inset; the scaling factor for the
  X--ray light curve (flux integrated in the 0.3--10 keV, see
  {\it 7}) is also shown. The X--ray light curve also includes two
  MAXI data points at t~=~3257~s and t~=~8821~s (empty squares). The
  fit performed over all the light curves required 24 free parameters
  (curve normalizations, host galaxy optical flux in each band, three
  temporal slopes and two breaks). Because of short-term low-level
  variability superposed to the long-term behavior and possibly
  residual inhomogeneity of optical data taken from different
  telescopes, we added in quadrature a 9\% systematic error in the
  optical and 5\% in the X--rays. Final fit yielded $\chi^2$ = 543.02
  / 565 d.o.f. A contribution from the host galaxy has been taken into
  account in the optical bands by fitting a constant flux of
  $\sim$~0.01~mJy for the reddest bands, corresponding to
  r$_{\rm{HG}}$~=~21.26~mag as tabulated in the SDSS catalogue. As a
  test of consistency, we performed the fit using a broken power law
  for the $B$, $g'$, $r'$ and $i'$ filters individually.
  Uncertainties are larger, but we do find that the values of $t_{\rm
    break}$ as well as the decay indices before and after the break
  time are still consistent, within the errors, with the values
  obtained by the overall fit.}

\newpage 

\begin{figure}[hbtp]
\begin{center}
\includegraphics[width=16cm,]{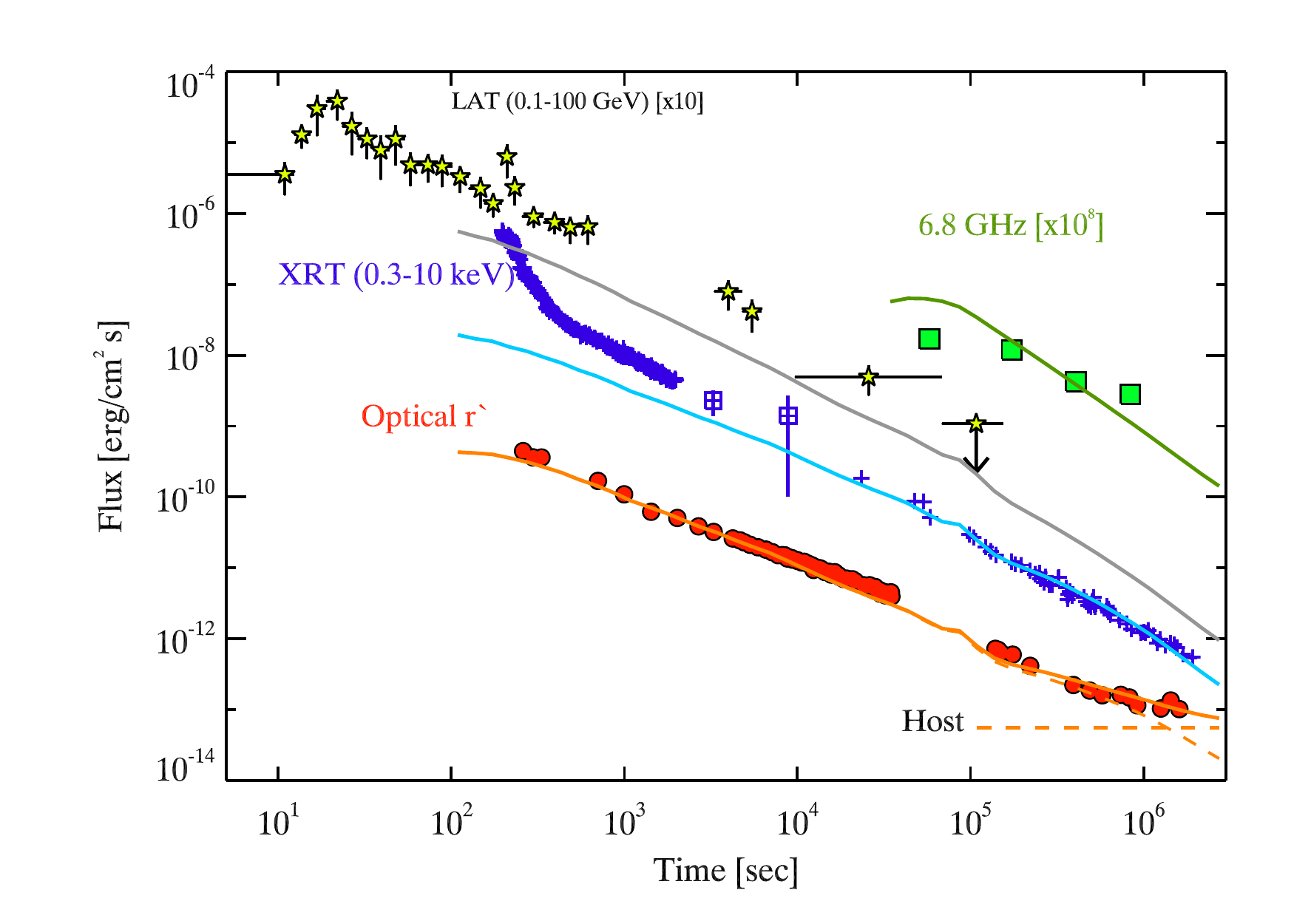}
\end{center}
\end{figure}
\vskip -1 cm
\noindent{\footnotesize{\bf Figure~3.} Radio (from {\it 31}: all measurements are taken at 6.8~GHz
  but the later one, at 7.3~GHz), optical, X--ray (Figure~2) and LAT
  $\gamma$-ray (from {\it 2}) light curves of \grb\ and
  corresponding model predictions adopting a description in terms of
  the van Eerten et al. model ({\it 25}). To properly
  fit the radio data, only a fraction of the electrons must be
  accelerated after $\sim 70$~ks (see Table~S10 and discussion in {\it 7}).}

\newpage 

\begin{figure}[hbtp]
\begin{center}
\vskip -3 cm
\includegraphics[width=14.5cm,]{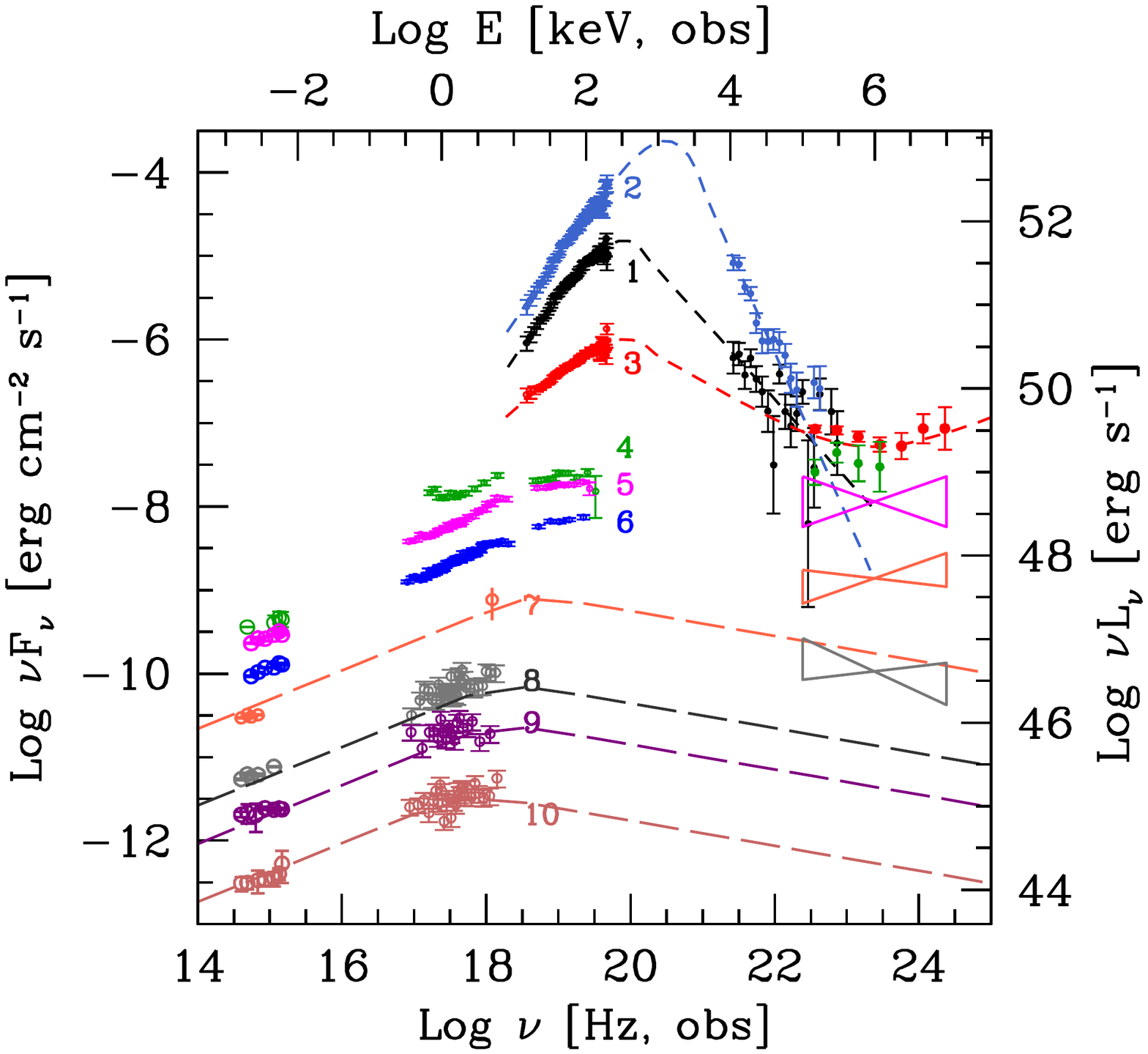}
\end{center}
\end{figure}
\vskip -5 cm
\noindent{\footnotesize{\bf Figure~4.} Spectral Energy Distributions
  (SEDs) of \grb\ taken at different times, from the optical to the
  GeV bands (LAT data from {\it 2}). For SED~1 and SED~2 the model is
  a Band function, for SED~3 the model is a Band+power law. Short
  dashed lines: phenomenological fit to the BAT+LAT data. Long dashed
  lines: results of the van Eerten et al. model ({\it 25}), with
  parameters discussed in ({\it 7}). The different SEDs refer to the
  following time intervals: 1: [0--6.5~s]; 2: [7.5--8.5~s]; 3:
  [8.5--196~s]; 4: [352--403~s]; 5: [406--722~s]; 6: [722--1830~s]; 7:
  around 3~ks; 8: around 23~ks; 9: around 59~ks; 10: around 220~ks
  (host galaxy contribution has been subtracted in this case). Note
  that from the optical SED analysis the intrinsic extinction is
  negligible.}

\newpage

\section*{\normalsize{Supplementary Materials}}

\noindent{www.sciencemag.org}\\
\noindent{Materials and Methods}\\
\noindent{Figures S1, S2, S3, S4, S5, S6, S7}\\
\noindent{Tables S1, S2, S3, S4, S5, S6, S7, S8, S9}\\
\noindent{References (32--62) [Note: The numbers refer to any
    additional references cited only within the Supplementary
    Materials].}

\newpage

\section{Observations and Data Analysis}

\subsection{\sw\ Discovery and Observations}
\label{sect:01:1}

\subsubsection{BAT Observations}
\label{sect:01:1:1}

BAT triggered on \grb\ ({\it 1}) at 2013-04-27 07:47:57.5~UTC (trigger
\# 554620) 51.1 seconds later than the \frm-GBM trigger ({\it 32}),
after \sw\ completed a pre-planned slew.
Hereafter, all times are referred to \tg\ so that t~=~T$-$\tg.
The BAT trigger was on the tail of the main peak, in a 64-second image
trigger (see ({\it 4}) for details on the BAT triggering system).
The \sw\ slew to the source started at t~=~148~s and ended at t~=~192~s.
The BAT position of the burst, initially calculated onboard and then
refined in ground analysis, is \ra\ = 11\h\ 32\m\ 36\s.1, \dec\ =
+27\D\ 42\am\ 20\as.3 with an uncertainty of 1\am.0 (radius, sys+stat,
90\% containment); this is 49\as.9 from the radio afterglow position
\ra\ = 11\h\ 32\m\ 32\s.82, \dec\ = +27\D\ 41\am\ 56\as.06 determined
with a precision of 0\as.4 ({\it 33}).

The BAT mask--weighted light curve in the 15--350~keV band
is shown in Figure~S1.
A first pulse beginning at t~=~0.13~s and peaking at t~=~0.5~s is
followed by another, smaller pulse at t~=~1.1~s.
Then the main episode of emission begins gradually at t~=~2.2~s,
with a sharp pulse at t~=~5.4~s.
A multi-peaked intense emission follows, lasting a total of about 5
seconds with a peak at t~$\sim 8$~s. 
A few, less intense pulses follow on top of a decay from the main
episode, with rise and decay on a time scale of a few seconds, the
last peaking at about t~=~26~s.
There was some significant deadtime over the main peak of the event,
that was corrected with a maximum correction factor of 1.72.
At t~=~120~s we have a fast rise with two overlapping pulses peaking at
t~=~131~s and t~=~141~s, respectively, followed by an exponential decay. 
After t~=~270~s there are no further prominent features. 
The decaying emission in the BAT energy range, well fit by a power-law
model, was detectable until the end of the observation at t~=~2021~s.
BAT data corresponding to the last 1031~s were collected in survey
mode.

\subsubsection{XRT Observations} 
\label{sect:01:1:2}

XRT data were accumulated in Windowed Timing (WT) and Photon Counting
(PC) mode ({\it 6}) depending on the brightness of
the source.
The pointed \sw-XRT observations of \grb\ started at t~=~195~s in WT~mode.
Due to the loss of star tracker lock that occurred soon after the
beginning of the observation, the attitude file needed for data
reduction has been manually reconstructed using the UVOT data (see
section~1.1.3) to provide time-dependent pointing corrections for the
XRT data.
For the subsequent observations the loss of star tracker lock occurred
again for several further short time intervals: data affected by bad
attitude reconstruction were adequately screened during the data
reduction process.
\sw-XRT observations that were used for this work include a total net
exposure time of $\sim$1826~s in WT~mode and $\sim$~203~ks in PC~mode
up to the sequence 090, spread over a $\sim$~4~Ms baseline.
The XRT data set was first processed with the XRTDAS software package
(v.2.8.0) developed at the ASI Science Data Center (ASDC) and
distributed by HEASARC within the HEASoft package (v.~6.13). 
Event files were calibrated and cleaned with standard filtering
criteria with the {\sc xrtpipeline} task using the latest calibration
files available in the \sw\ CALDB.
Standard grade filtering was applied: 0--12 for PC data and 0--2 for
WT data.
The list of all XRT observations of \grb\ used for the present
analysis is shown in Table~S1.

The first XRT observation in WT mode partially overlaps the time
interval during which BAT data are available, from its start until
t~=~990~s; the remaining part is given by 1031~s of data.
The count rate of the first XRT observation was high enough to cause
severe pile-up in the WT mode data.
To account for this effect, the WT data were extracted in annular
regions with a 60-pixel (1 pixel = 2.36\as) external radius and a
variable radius of the inner excluded region depending on the pile-up
degree.
The size of the exclusion region was determined following the
procedure illustrated in ({\it 34}).
The values adopted for the inner radius are reported in Table~S2.

Data from light curves and spectra relevant to observations in PC mode were
extracted using the task {\sc xrtgrblc}, that performs the appropriate
corrections for vignetting and PSF losses.
The task optimizes the extraction regions according to the count rate
of the source in each orbit, excluding a central circular region in
case of pile-up (sequences 001 to 013) whose radius is reported in
Table~S2.
The background was extracted from an annular region with inner and
outer radii 47 and 90 pixels, respectively.
Ancillary response files were generated with the {\sc xrtmkarf} task
allowing us to correct for CCD defects.
We used the latest CALDB response matrices (rmf): v013 (PC mode) and
v014 (WT mode).

\subsubsection{UVOT Observations}
\label{sect:01:1:3}

\sw\ Ultraviolet Optical Telescope (UVOT; ({\it 5,35,36})) began
observing the burst at t~=~181~s.
During the automatic sequence, \sw\ was unable to obtain a positional
lock using the onboard catalogue as there were too few guide stars in
the field of view.
The loss of star tracker lock caused the optical afterglow to wander
across the image plane, so that with each consecutive exposure the
afterglow position shifted increasingly in RA and Dec away from the
reported position.
For longer exposures this movement caused the point sources in the
image to be trailed.
Since these observations were taken in event mode (where the position
and timing of each event are recorded down to a time resolution of
11~ms), we were able to aspect correct the observations to a finer
time resolution than the duration of each exposure and recover the
trailed sources as point sources.
Because the observed and actual positions of the stars in the image
were different by up to $\sim$~300\as, the aspect correction was
performed in a two-step process.
First, a coarse aspect correction was determined manually for each
exposure; then, a fine aspect correction was determined by extracting
an image from the event list every 10~s and cross-correlating the
stars in the image with those in the USNO-B1 catalogue.
The differences in RA and Dec were then applied to each event falling
within the particular 10~s interval ({\it 37}). 

During the first 2~ks of observations, the optical afterglow of
\grb\ was so bright that UVOT suffered from heavy coincidence losses
and scattered light.
The $v$-band 10~s settling point observed from t~=~181~s, the majority
of white data before T+2000~s, and the $b$ and $u$ band data observed
before 650~s could not be recovered.
However, using the read-out streaks ({\it 38}) we were
able to obtain photometry for the white band data between 506 and
1023~s and for the first $b$ and $u$ band exposures.
Earlier and later saturated white data are beyond the recommended
range for the read-out streak method.
We were also able to obtain photometry from UVOT UV grism spectrum,
which provides the earliest UVOT photometry for this GRB, by folding
the spectrum through the filter response curves.

For the later follow-up observations, the star tracker obtained a
positional lock on most images, but for the observations where it
failed to lock the images are trailed or the point sources are
blurred.
Therefore, we manually inspected all exposures and excluded those that
would produce unreliable photometry.
For the aspect corrected event mode data, the photometry was obtained
using the uvot tool {\sc uvotevtlc}, while the image mode data were
processed using the uvot tool {\sc uvotmaghist}.
The photometry for the image and event mode observations were
extracted using a circular aperture with a radius of 5\as\ when the
count rate was above 0.5 \cts, and 3\as\ aperture when the count rate
had dropped below 0.5 \cts; an appropriate background region was used.
We applied coincidence loss corrections and standard photometric
calibrations ({\it 39-41}).
The analysis pipeline used software HEASOFT~6.13 and UVOT
calibration 20130118.
The list of UVOT observations which were included in the analysis of
\grb\ are shown in Table~S3.

Since November 2008 the automated sequence that is triggered by a
strong burst includes a 50-second UV grism exposure after the first
white finder chart.
A magnitude brighter than $v$~=~12~mag is needed to get a GRB
detection.
\grb\ provided the second good early UV spectrum from the instrument.

The data reduction required special adaptations due to the loss of
star tracker lock which affected the normal processing.
Using the data of detected photon positions in the white filter prior
to the grism exposure in image mode, and assuming a steady motion, the
spacecraft moved during the grism exposure by about 9.5\as (18 pixels)
under an angle of 64$^{\circ}$ to the dispersion in the direction of
longer wavelengths.
Assuming a smooth motion, this translates into a shift of
$\approx$~54~\AA\ in the dispersion direction.

The blurring effect of the motion on the grism image was determined by
deconvolving a zeroth order of a fainter object in the grism image by
a zeroth order from an image obtained during normal \sw\ operations,
i.e., with attitude lock.
The resulting kernel shows the exposure locations on the detector for
that zeroth order and thus suggests that the exposure was mainly taken
whilst the pointing rested in two locations about 7 pixels apart under
an angle consistent with the analysis from the white filter event
data.
Using that kernel to deconvolve the grism image, a cleaner grism image
was obtained where the effects of blurring are to a large extent
removed.  
For the deconvolution the STARLINK LUCY program was used.

To determine the position of the spectrum on the detector, the last 5
seconds of the event data from the white filter finder were separately
processed and aspect-corrected using the HEASOFT~6.13 tools.
This image ended 5.5~s before the start of the grism exposure,
so the position of the source on the detector in the white filter was
known.
Using the new grism calibration and related software
(http://www.mssl.ucl.ac.uk/www\_astro/uvot/) this was used to
determine the location of the spectrum on the grism image and extract
the spectrum with good knowledge of the wavelength scale.

The flux calibration used includes a correction for coincidence loss
which is estimated to be in the range of 10-20\% for this spectrum.
There is partial overlap of second order emission, estimated to start
affecting the flux above the observed wavelength of 3000~\AA ~by 15\%,
slowly varying thereafter.
The spectrum observed above 3000~\AA\ can be used to some extent for
the spectral lines present, for example, redshifted Mg~II
2800~\AA\ resonance line is present as one of the strongest
absorptions.

The \sw\ UVOT spectrum with wavelengths converted to the rest frame of
the GRB host is shown in Figure~S2.
The wavelength scale in the UVOT grisms is not calibrated on board and
its accuracy relies on the knowledge of the position of the source.
A Mg~II 2800~\AA\ resonance line seems present with a significance
larger than 4~$\sigma$, although not exactly at the right frequency
location.
The width of this line is larger (by a factor of two) than expected
from the estimated satellite motion from the deconvolution kernel,
which would be of order of 15~\AA\ only if not already removed by the
image deconvolution.

\subsection{MAXI Observations}
\label{sect:01:2}

The Gas Slit Camera (GSC; ({\it 42})) of Monitor of
All-sky X--ray Image (MAXI; ({\it 15})) detected
\grb\ with one of its cameras (camera~\#4) at the two consecutive scan
transits centered at t~=~3257~s and t~=~8821~s.
The effective area of the GSC camera during the transit had a
triangular shape with the FWZM duration of 52~seconds and the peak
effective area of 3.9~cm$^2$ at these observations.
In the present analysis, we assumed the flux within a transit to be
constant.
This assumption is justified considering that the time interval since
the trigger ($\sim$ thousands of seconds) is much longer than the
transit time ($\sim$~50~s), and therefore the variation within the
transit is negligible.
MAXI usually scans a specific position in the sky once every International 
Space Station orbit ($\approx$~92~min period).
No significant flux was detected in the GSC Cameras from the source at
the last scan transit of the GRB location before the trigger ($t =
-2308$~s) and at the third scan after the trigger ($t = +14385$~s).
The calibration of the energy response and details of its flight
performance is described in ({\it 43}).

\subsection{BAT and XRT Spectral Analysis}
\label{sect:01:3}

We extracted several 15--150 keV BAT spectra at different time
intervals starting from t = $-$0.1~s up to the end of the BAT
observation. 
We fitted all these spectra with a simple power-law model and the
best-fit parameters are reported in Table~S4.
A significant spectral evolution is apparent in the spectral results:
the variation of the power-law photon index is shown in the bottom
panel of Figure~S3.

In the time interval between t~=~195~s and t~=~990~s, corresponding to
the overlap between the BAT data and the XRT-WT data, we carried out a
simultaneous broad band (0.3--150 keV) spectral analysis, building the
BAT and XRT spectra in nine common time intervals (see Table~S4).
After the end of the BAT observation, the spectral analysis is based
only on the 0.3--10 keV XRT data.
Both the WT spectra collected between t~=~990~s and t~=~2021~s and the PC
spectra collected after t~=~$2.2\times 10^4$ s can be well described by
an absorbed power law with a column density fixed to the line-of-sight
value of $\rm 1.8 \times10^{20} cm^{-2}$ plus an intrinsic (redshifted) 
and time varying (in the first 1~ks) absorption (see Table~S4).

We are aware that fitting a slightly curved spectra with a power law
introduces small biases in the column density estimate (e.g. ({\it
  44})).
Therefore, we do not consider the column density increase derived from
power-law fitting of the afterglow data to be physically significant.

We have verified that no significant spectral variations occur during
the XRT-PC monitoring, with the power-law index consistent with a
constant value.
Thus we have obtained an average PC spectrum from observation
sequences 001 to 039: the results of the averaged spectral analysis
are reported in Table~S4.

\subsection{Flux-calibrated X--ray Light Curve}
\label{sect:01:4}

We used the data from \sw-BAT, \sw-XRT and MAXI/GSC to build a
flux-calibrated X--ray light curve in the 0.3--10~keV band.
We used the results of the BAT spectral analysis in the 15--150~keV
band to compute the conversion factor needed to obtain the
0.3$-$10~keV flux from the count rate of the 64~ms BAT mask-weighted
light curve up to t~$=$~202~s.
This conversion factor was computed with higher precision in the
time interval in which the BAT and XRT data overlap using the results
of the simultaneous BAT and XRT spectral analysis.
The 0.3--10~keV flux for the MAXI/GSC data was obtained assuming the
power-law model with a photon index $\Gamma_M$ = 1.8~$\pm$~0.1 derived
from the spectral analysis carried out on XRT data at close epochs.

The 0.3$-$10~keV flux-calibrated light curve that we obtained is shown
in Figure~S3.
The good agreement between the BAT and the XRT light curves when they
overlap can be interpreted as the sign that the early X--ray light
curve (t $< 10^3$~s) is probably still related to the prompt emission.
For $t~>~260$ s the X--ray light curve can be fitted with a
double-broken power law.
The first decay is steep with $\alpha_{0, \rm x} = 3.32\pm0.17$ and a
break at $t_{1, \rm x}~=~424~\pm~8$~s.
The decay then flattens to $\alpha_{1, \rm x}~=~1.28~\pm~0.01$.
A further break is needed, even if not well constrained.
The second break occurs at $t_{2, \rm x}~=~48~\pm~22$~ks and the
slope steepens to $\alpha_{2, \rm x}~=~1.35~\pm~0.02$.
We remark that it is only thanks to the very good quality of our data
that we were able to reveal this further break, whose significance is
$3.8~\sigma$, estimated through an F-test.
The reported errors are derived computing the interval for a $\Delta
\chi^2=2.71$ and are therefore the errors at $90\%$ confidence level
for one parameter of interest.
A $5\%$ error has been summed in quadrature to all the points to
obtain a good fit with a reduced $\chi^2_{\rm red}=0.9$ with 243
degrees of freedom.
These decay indices and break times lie well within the distribution
of those parameters as seen for the GRB population as a whole ({\it
  45,46}) \footnote{see http://www.swift.ac.uk/xrt\_live\_cat for
  update versions of those populations}.

We plot in Figure~S4 all the X--ray light curves of the
GRBs observed by \sw-XRT (627 GRBs as of June~4,~2013).
As it can be seen, \grb\ has the brightest X--ray afterglow, in term of
observed flux, that have ever been observed by XRT.
But this is due both to the intrinsic brightness of this burst and to
its proximity that enhance its fluxes.
A more informative plot is the comparison of the luminosity of this GRB
with those of other bright GRBs.
To this end we plot in Figure~S5 the rest-frame
isotropic luminosity curves for a sub-sample of GRBs observed by \sw\ (both
X--ray and $\gamma$-ray luminosity).
We do not show the faintest GRBs, to avoid overcrowding, and include
also the brightest Fermi GRBs.
As can be seen, even if \grb\ has the brightest X--ray afterglow in
term of observed flux, it has a normal behavior in term of rest-frame
luminosity.
In fact, although it lies at the bright end, it is fully consistent
with the range displayed by the long GRBs population at large.

\subsection{Ground--based Optical/NIR Observations}
\label{sect:01:5}

The 2--m Faulkes Telescope North (FTN) robotically followed up
\grb\ starting at t = 5.1~minutes, corresponding to $\sim$ 220~s in
the rest frame.
Observations were carried out with a scheduled sequence of images with
the $B$, $V$, $r'$, and $i'$ filters ({\it 47}) for
the subsequent 5~hours. 
From 2.4~hrs after the detection, observations were performed also
with the MITSuME Telescopes (0.5-m Akeno Observatory, 0.5-m Okayama
Observatory and 1.05-m Ishigakijima Observatory; ({\it 48})) using 
the $g'$, $R_{\rm c}$, and $I_{\rm c}$ 
filters and following the optical afterglow up to 5.2~days.
Data in these filters were calibrated with respect to the SDSS $g'$,
$r'$, and $i'$ bands, respectively.
Late time observations have been performed also with the 2-m Liverpool
Telescope (LT) in the time interval between 1.5~days and $\sim$~19~days
after the burst, using the Sloan filters SDSS-$r'$ and SDSS-$i'$.

Calibration of the entire optical data set was carried out with
respect to a common set of selected field stars.
Sloan Digital Sky Survey (SDSS) catalogued stars were used for the
$g'$, $r'$ and $i'$ filters, while standard stars were used for the
$B$ and $V$ filters.
A summary of our calibrated data, with a log of all the observations
and magnitudes of the optical counterpart in all filters, can be found
in Tables S5--S9.

\subsection{Consistency with Spectral Energy Correlations}

With the exception of very few outliers, the long GRBs obey the
spectral energy correlations between the rest-frame peak energy
$E_{\rm peak}$ of their spectrum and the overall isotropically
equivalent energetics $E_{\rm iso}$ ({\it 28}) or
isotropic peak luminosity $L_{\rm iso}$ ({\it 29}).
It is interesting to verify whether \grb\ follows these correlations as
well.
Figure~S6 shows how \grb\ is located in both the ``Amati"
and the ``Yonetoku" planes.
As can be seen, it lies exactly on the best fit of both.
In the same figure, the green stars show GRBs associated with a
spectroscopically confirmed supernova ({\it 14}).
Note how \grb\ stands out.

If the break at $\sim$~37~ks is indeed a jet break, we can calculate the
opening angle of the jet $\theta_{\rm j}$.
With a homogeneous circumburst medium of density $n=1$ cm$^{-3}$ and
a kinetic energy of the fireball equal to $E_{\rm k, iso}=4\times 10^{54}$
erg, we derive $\theta_{\rm j}\sim3^\circ$.
The corresponding collimation corrected energy is $E_\gamma=10^{51}$ erg.
This makes \grb\ consistent (at 2$\sigma$) also with the correlation
between $E_{\rm peak}$ and the collimation corrected energetics $E_\gamma$
(the so-called ``Ghirlanda" correlation, see ({\it 30})).

\newpage

\section{Result of the SED modeling}
\label{sect:02}

In this section we will try to establish if the extraordinary richness
and quality of data of GRB 130427A can be explained in the simple and
standard framework of a relativistic fireball running into the
circumburst medium (CBM), that decelerates the fireball and produces
two shocks: the forward shock, running into the CBM, and the reverse
shock, running into the ejected material of the fireball.
To this aim we adopt a publicly available model, developed by ({\it
  25}), that captures with some details the main physical processes we
believe are occurring.
On the other hand, reality is surely more complex than a simple
standard model can account for.
There can be more specific versions of the standard model, e.g. the
fireball could be structured (namely, having an angular profile of
bulk Lorentz factors and energies), the CBM could be clumped, the
microphysical parameters could be varying in time and in space.
However, all these are ``versions" of the standard model (a collimated
fireball running into the CBM) that we aim to test in its pillars, and
not in its ``details".
We therefore ask: does the basic model reasonably account for the
observed properties of this data--rich burst, or should we drastically
reconsider our current general view of GRB physics?

\subsection{Main assumptions}
\label{sect:02:1}

The similarity of the X--ray and optical light curves after
$\sim~10^4$ seconds suggests that they are produced by the same
process.
The similarity of their spectral indices indicates that they belong to
the same power-law spectrum.
Therefore, we assume that the optical and the X--ray emission, after
$\sim 10^4$ seconds, are both afterglow emission produced by a forward
shock.
Before $10^4$~s, an extra component contributes to the emission, most
notably in the X--rays.
This extra component could be late prompt emission or reverse shock
emission, but we will not model it.

The \frm--LAT emission, above 100~MeV, is not correlated with the
$\sim$~MeV emission as shown by the GBM or BAT, but shows a peak at
$\sim$~20 seconds followed by a power-law decay (see ({\it 2})).
These properties suggest that the \frm-LAT flux is afterglow emission
produced by a forward shock.
The peak of the emission indicates the onset of the afterglow.
We also assume that the LAT luminosity above 100~MeV is a proxy of the
bolometric luminosity of the afterglow.
This should be truly independent of the radiation process originating
this high-energy emission.

We assume that the radio emission could either be the low frequency
part of the X--ray and optical forward shock emission, or instead be
due to a reverse shock as suggested by ({\it 31}).
In the latter case our afterglow model, with no reverse shock
emission, should underestimate the observed radio emission.

We assume that the achromatic break in the light curves at $\sim$~37~ks
is a jet break.

\subsection{General derived properties}
\label{sect:02:2}

{\it Density profile ---}
In the case of a homogeneous density, $L_{\rm bol}$ increases during
the coasting phase ($\Gamma$=constant) as the observed surface area,
hence as $t^2$.
If the circumburst medium has a wind density profile $n\propto
R^{-2}$, the increase of the observed surface area in the coasting
phase is exactly compensated by the decreasing density, and the
bolometric luminosity is constant (see, e.g., ({\it 49}) and Eq.~11 in
({\it 50})).
After the deceleration time, for an adiabatic fireball we have $L_{\rm
  bol}\propto t^{-1}$ both for the homogeneous and the wind density
profile, as long as the emission occurs in the fast cooling regime.
This is independent from the radiation process producing the LAT flux
[e.g. synchrotron or synchrotron self--Compton (SSC)].

There exists the possibility that the GeV peak is produced by a
reverse shock, possibly connected with the early optical flux
({\it 10}). 
However, in this case, the decay slope of the optical flux depends on
the density profile, and if it is wind--like, it is steeper than
observed ({\it 51}).

The flux decay after the peak is suggestive of an homogeneous medium
if the GeV emission is SSC, since a wind density profile would make the
GeV light curve substantially steeper than observed (this point is made
very explicitly by ({\it 26})).
If the GeV is synchrotron, instead, we face the problem of explaining
the nature of the very energetic photons (up to 130~GeV, rest frame)
that are not expected in a shock acceleration scenario (see discussion
in ({\it 2})).
Another problem, shown in Fig.~11 of ({\it 52}), is the
over--prediction of the flux at early times ($t<300$ s) in the \sw-BAT
and \frm-LAT energy ranges.

A wind medium has been assumed to reproduce the X--ray/optical/radio
data of \grb\ ({\it 31,52}) and to explain the early optical flux
({\it 10}).
In ({\it 31}), the wind like density is preferred by comparing the
predicted and observed X--ray decay slope.
The slope predicted by the homogeneous density is {\it closer} to the
observed decay.
Nevertheless, they chose the wind density profile (compare the
following values derived by ({\it 31}): predicted decay in the wind
case: $\alpha_x=1.64$, in the homogeneous case: $\alpha_x=1.14$,
observed value: $\alpha_x =1.35$).
Furthermore, their wind afterglow model requires a kinetic energy
$E_{\rm k,iso}~\sim~10^{52}$~erg, which is only 1\% of the prompt
emission energy ($E_{\rm \gamma,iso}~=~10^{54}$~erg), which seems
rather low.
This small value of $E_{\rm k,iso}$ implies rather large values of the
microphysical parameters: $\epsilon_{\rm B}$ and $\epsilon_{\rm e}$
are of order unity.
As a consequence, the $\epsilon_{\rm e}/\epsilon_{\rm B}$ ratio is
also of order unity, and this limits the possibility to explain the
GeV emission as due to the SSC process.
It is then difficult to explain the detection of photons up to 130~GeV
(rest frame).
The normalization of the wind density profile is roughly 1,000 times
smaller than expected by a typical star wind, even if this parameter
is still largely uncertain for the (unknown) final phases of the life
of a massive star.

Finally, we have considered a generic stratified medium, with a
density $n\propto r^{-s}$, with $s=1$ and $s=1.5$, following ({\it
  53}), with the hope to ameliorate the problems of the pure
wind--like scenario.
We did find that these solutions present a deceleration peak flux,
that can explain the peaked GeV emission, but are worst
representations of the data at the other frequencies.

To conclude, i) the presence of the peak of the light curve and ii)
the relatively flat decay slope of the light curves at all frequencies
are observational evidences supporting the homogeneous CBM.
Further support of this hypothesis is given by the small kinetic
energy (requiring very large efficiencies) and the small wind
normalization required by the proposed wind models.
The latter are certainly possible, but the required (not mainstream)
assumptions and the remaining open issues, in our opinion, are more
than in the homogeneous case, that therefore we choose.

\vskip 0.2 cm
{\it Bulk Lorentz factor ---}
The time of the onset of the afterglow, for an assumed radiative
efficiency $\eta$ (defined through $E_{\rm iso}=\eta E_{\rm k, iso}$,
where $E_{\rm k, iso}$ is the isotropic equivalent kinetic energy of
the fireball) and ISM density $n$, allows us to infer the value of the
bulk Lorentz factor $\Gamma_0$ of the fireball before deceleration,
i.e. during the coasting phase (see the discussion
in ({\it 24}) about the existing formulae).
For $E_{\rm \gamma, iso}=10^{54}$ erg s$^{-1}$, $\eta=0.1$, $t_{\rm pk}=20$ s  
and a density $n=1$ cm$^{-3}$ we obtain
$\Gamma_0=505$ (using the formula in ({\it 24})), scaling
as $\Gamma_0\propto [E_{\rm \gamma, iso} /( n \eta t^3_{\rm pk})]^{1/8}$.

\vskip 0.2 cm
{\it Microphysical parameters: $\epsilon_{\rm B}$ ---} 
The fraction of the available energy at the relativistic shock going
to relativistic electrons and to magnetic energy is parameterized by
$\epsilon_{\rm e}$ and $\epsilon_{\rm B}$, respectively.
The fact that the X--ray and the optical light curves share the same
temporal profile in the $10^4$--$ 10^6$ s interval implies that they
belong to the same spectral power law branch.
Therefore the characteristic frequencies, such as the injection
frequency $\nu_{\rm inj}$ or the cooling frequencies $\nu_{\rm cool}$,
that are time--dependent, must be located outside the optical--X--ray
frequency range in this time interval because, otherwise, the light
curves could not be parallel.
The requirement $\nu_{\rm cool}~\gsim$~10~keV can be fulfilled if
$\epsilon_{\rm B}$ is small (of the order of $10^{-5}$--$10^{-4}$)
implying a modest cooling (in turn implying a large $\nu_{\rm cool}$).
This small value of $\epsilon_{\rm B}$ is not unprecedented, since it
agrees with those found by ({\it 54}) for the GRBs
detected in the GeV band by the LAT on board \frm.
However, more often $\epsilon_{\rm B}$ is around $\sim
10^{-2}-10^{-3}$ ({\it 19,53}).

\vskip 0.2 cm
{\it Microphysical parameters: $\epsilon_{\rm e}$ ---} 
If the LAT emission is afterglow by a forward shock, and if it is a
good proxy for the bolometric luminosity ($L_{\rm LAT} \, = f \,
L_{\rm bol}$), we can derive $\epsilon_{\rm e}$.
In the fast cooling regime, $L_{\rm LAT}$ after the deceleration time
is given by (see e.g. ({\it 24}))
%
\begin{equation}
L_{\rm LAT} \, = f \, L_{\rm bol} \, =\, 
{3 f\epsilon_{\rm e}\over 2} \, {  E_{\rm kin,iso} \over t};  
 \quad t \ge t_{\rm pk}; 
\qquad
E_{\rm kin, iso} \, =\, E_{\rm \gamma,iso}\, \left( {1-\eta \over \eta} \right)
\end{equation}
therefore
\begin{equation}
\epsilon_{\rm e}\, \sim \, {2  \over 3f  }{\eta\over 1-\eta } 
{L_{\rm LAT} \over  E_{\rm \gamma, iso}}  t_{\rm pk} 
\end{equation}
With $E_{\rm \gamma, iso}\sim 10^{54}$ erg, $\eta=0.1$, $f\sim 1/6$
and $L_{\rm LAT}\sim 1.6 \times 10^{51}$ erg s$^{-1}$ at $t_{\rm
  pk}\sim 15$ s (in the rest frame), we derive $\epsilon_{\rm e}\sim 10^{-2}$.
Notice that {\it this is independent of the emission process} (see the
discussion after Eq.~1 in ({\it 55})).

\vskip 0.2 cm
{\it Jet opening angle and viewing angle ---}
Defining the viewing angle $\theta_{\rm v}$ as the angle between the
line of sight and the jet axis, we have that a non--zero $\theta_{\rm
  v}$ produces two breaks in the light curve, the first when $1/\Gamma
= \theta_{\rm j}-\theta_{\rm v}$, and the second when $1/\Gamma =
\theta_{\rm j}+\theta_{\rm v}$.
With respect to an observer located on the jet axis ($\theta_{\rm
  v}=0$) the off--axis viewer should see the jet break as a smoother
transition (e.g. ({\it 25,56})).
This could help make the post--break light curve decay less steeply
than predicted, as in our case, up to the second break.
The latter could be hidden in the optical by the host galaxy+supernova
contribution to the light curve flux, but should be visible in the
X--ray light curve.
Note that a shallow break can also be the result of time-varying
microphysical parameters: if $\epsilon_{\rm B}$ and $\epsilon_{\rm e}$
increase, then the light curve decays less steeply than predicted by
the simplest theory, at least up to the time when $\epsilon_{\rm B}$
and $\epsilon_{\rm e}$ stop increasing (since the range of their
values is limited).
After that time, the light curve should decay as expected by the simple
theory.

\vskip 0.2 cm
{\it Radio spectra ---}
The available radio data (as in ({\it 31})) indicate a
rather flat spectrum, inconsistent with self--absorption and roughly
consistent with $F_\nu \propto \nu^{1/3}$, the single electron
synchrotron spectrum.
This indicates that in the radio range $\nu_{\rm a}< \nu<\nu_{\rm inj}$.
A small $\nu_{\rm a}$ in turn requires a small density (of the
relativistic electrons responsible for the synchrotron emission and
absorption), much smaller than the typical value of $n\sim 1$ cm$^{-3}$.
A large $\nu_{\rm inj}\propto \gamma^2_{\rm inj} B\Gamma$ requires
large energies of the electrons.

\subsection{Modeling}
\label{sect:02:3}

\vskip 0.2 cm
{\it The model ---} 
To model the SEDs and the light curves at any frequency of \grb\ we
use the model developed by ({\it 25}).
This model considers the synchrotron emission produced by the forward
shock of an adiabatic fireball expanding into a homogeneous medium.
The fireball is canonical (namely, matter, not magnetically dominated,
and uniform, i.e. with radial velocity directions that are not a
function of the angle from the jet axis) with an opening angle
$\theta_{\rm j}$, viewed at an angle $\theta_{\rm v}$ from the axis.
It accounts for the so called ``jet--break", i.e the break in the
light curve occurring when $1/\Gamma\sim \theta_{\rm j}$.
Arrival times from the emitting volume are properly calculated.  The
model does not include the emission from the reverse shock, nor the
contribution from synchrotron self--Compton (SSC) process.
The lack of SSC is a rather serious limitation in the first (fast
cooling) phases of the afterglow if the Comptonization parameter $y$
(regulating the importance of the SSC radiation) is large, while in
slow cooling the importance of SSC radiation rapidly decreases.
Although we require a small $\epsilon_{\rm B}$, and $y \sim
(\epsilon_{\rm e}/\epsilon_{\rm B})^{1/2}$ (see above), our modeling
pertains to the slow cooling phase, therefore the lack of SSC does not
affect our results.
Table~S10 lists the parameters of the models shown in Figure~S7 (light
curves) and Figure~3 of the main journal article (SEDs).

\vskip 0.2 cm
{\it The LAT light curve ---}
As mentioned above, the time profile of the bolometric light curve in
fast cooling is independent of the emission process.
If the LAT flux is indeed close to the bolometric flux, then its
behavior cannot be used to discriminate between -- say -- synchrotron
and SSC emission, provided that both processes are capable of
generating high-energy photons.
For \grb\ photons exceeding 40~GeV have been observed, with one
exceeding 90~GeV, and this represents a problem for the synchrotron
interpretation of the high-energy flux (see, e.g. ({\it 57})).
On the other hand, both SSC and inverse Compton scattering between the
accelerated electrons and the prompt emission photons still present in
the fireball can contribute to the GeV flux.
This is even more likely if $\epsilon_{\rm B}$ is indeed very small.
In this case the Comptonization $y$ parameter is bound to be large in
fast cooling, making the Compton process the dominant one (unless the
synchrotron typical frequencies, in the comoving frame, are so large
as to make the Compton process to be in the Klein--Nishina regime).

Ghisellini et al. ({\it 55}) discussed the case of a
radiative fireball to explain the time decay of several GRBs whose
$\gamma$--ray flux decayed as $t^{-1.5}$ even though their spectral
slope was close to $\Gamma_\gamma=2$.
The decay slope of \grb\ is instead close to unity, i.e. what is
expected from the bolometric luminosity decay of an adiabatic
fireball.
This can be explained if we consider that in \grb\ the afterglow onset
time is close to 15 seconds: this may appear to be a very short time,
but it is nevertheless later than the onset time of the bursts
considered in ({\it 55}).
This relatively late onset implies that the total energy emitted in
the LAT energy band is significantly lower than the prompt emission
energy.
If the prompt emission did not contain a sizable fraction of the
kinetic energy of the fireball, then 
$E_{\rm k}\sim E_{\rm \gamma, iso}/\eta$  and does not vary during 
the $\gamma$--ray emission.  
In other words, the fireball remains adiabatic or quasi--adiabatic.
We stress that if the ISM had a wind density profile, then the initial
light curve of the bolometric luminosity would be flat, without any peak.
Therefore an homogeneous medium is consistent with interpreting the
peak of the LAT light curve as the afterglow onset, while a wind
density profile would require another interpretation for the same
peak.

\vskip 0.2 cm
{\it The XRT light curve --- }
The model reproduces the X--ray light curve after $\sim10^4$ seconds.
We suggest that before this time the X--ray flux is dominated by late
prompt emission and high--latitude radiation.
By ``late prompt" emission we mean radiation produced by the same
mechanism(s) producing the main GRB event, but occurring later.
By ``high--latitude" radiation we mean the radiation from parts of the
emitting surface not on axis with the line of sight.
If the surface switches off abruptly, the distant observer will see
that a flux decaying in time at a predictable rate, as a result of the
different arrival time of the photons combined with the different
degree of beaming.
Indeed, the XRT and BAT light curves are very similar as long
as BAT detects the burst ($\sim$1900 seconds).

\vskip 0.2 cm {\it The optical--UV light curve --- } 
In Figure~S7 we show how the model reproduces the best
sampled optical light curves, in the $R, r^\prime$ and $R_{\rm c}$
filters.
In the other optical bands the behavior is the same. 
In the model of the optical emission shown in Figure~S7, we have added
a constant contribution corresponding to $R=21.3$ (flux density
$\sim$0.0113~mJy) due to the host galaxy (dashed line in Figure~S7).

\vskip 0.2 cm
{\it Jet break ---}
We interpret the break at $\sim 37$ ks as a jet break, corresponding to a jet 
opening angle of $\sim 3.4^\circ$.
We assume $\theta_{\rm v}=0^\circ$.
In fact, although the assumption of a non--zero $\theta_{\rm v}$ could
help to explain the shallow decay of the post--break light curve
($\propto t^{-1.3}$, shallower than predicted by the standard model,
e.g. ({\it 58})), the late X--ray data (not affected
by the host galaxy+supernova emission) do not show any further
steepening of the light curve up to t~$\sim$~45 days.
We are then led to assume that the radiation emitted in the
post--break phase is larger than what simple theory predicts (shown by
the dotted models in Figure~S7).
This can be the result of having energy injection, increasing the total
fireball energy which in turn increases the amount of emitted radiation.
Alternatively, we can maintain a constant fireball energy, but increase
the amount of energy given to the electrons (i.e. $\epsilon_{\rm e}$).
We prefer the latter solution because it has a lower energy requirement.
We then assume that $\epsilon_{\rm e} \propto (t/t_\epsilon)^{0.6}$
after $t_{\epsilon}=0.8$ day.
Then, to obtain a reasonable agreement between the optical and X--ray
data (circles and crosses in Figure~S7) and the model, we
also assume that $\epsilon_{\rm B} \propto (t/t_\epsilon)^{0.5}$ after
the same $t_{\epsilon}~=$~0.8~day.
Possible evolution of the microphysical parameters at the shocks have
been invoked to explain the multiwavelength light curves (in the
X--ray and optical bands) of some GRBs
(e.g. ({\it 59})).

\vskip 0.2 cm
{\it The radio light curve --- }
The model that we have described above, which assumes an evolution of
the $\epsilon_{\rm e} $ and $\epsilon_{\rm B}$ parameters,
over--predicts the radio flux (dot--dashed line vs squares in Figure~S7).
This is due to the fact that the $\nu_{\rm inj}$ frequency is too
small and it is even smaller than the self--absorption frequency
$\nu_{\rm a}$.
The injection frequency is $\nu_{\rm inj}\propto \gamma_{\rm
  inj}^2B\Gamma \propto \epsilon_{\rm e}^2 \epsilon_{\rm B}^{1/2}
t^{-3/2}$.
Changing the microphysical parameters inevitably leads to other
inconsistencies.
An extra degree of freedom is the fraction of the electrons that are
accelerated at the shock front and that produce the radiation received
by the observer.
Instead of assuming that the shock accelerates one electron per
proton, we can assume that only a fraction $k<1$ of the present
electrons are accelerated.
In this way we can increase the energy per electron and, therefore,
increase $\nu_{\rm inj}$.
This, however, must occur in such a way as to satisfy the requirement
that $\nu_{\rm inj}$ is smaller than the observed optical frequencies,
at least between $10^4$ and $10^6$ seconds, if the optical flux
remains a power law of the same slope in this time interval.

Admittedly, the complex behavior of the afterglow of \grb\ in the
radio, optical, X--ray and $\gamma$--ray bands cannot be fully
captured by a simple and standard afterglow model, e.g. synchrotron
and self Compton emission by a forward shock, with non-varying
microphysical parameters.
Therefore we must ask ourselves: does \grb\ require a complete
revision of our ideas about the origin of the afterglow emission?
Our answer is no: the unique richness of the \grb\ data can still be
explained within the framework of the standard model.
It surely requires extra assumptions, but not a radical change.

~~~~~~~~~~~~~~~~~~~~~~~~~~~~~~~~~~~~~~~~~~~~~~~~~~~~~~~~~~~~~~~~~~~~~~~~~~~~
~~~~~~~~~~~~~~~~~~~~~~~~~~~~~~~~~~~~~~~~~~~~~~~~~~~~~~~~~~~~~~~~~~~~~~~~~~~~
~~~~~~~~~~~~~~~~~~~~~~~~~~~~~~~~~~~~~~~~~~~~~~~~~~~~~~~~~~~~~~~~~~~~~~~~~~~~
~~~~~~~~~~~~~~~~~~~~~~~~~~~~~~~~~~~~~~~~~~~~~~~~~~~~~~~~~~~~~~~~~~~~~~~~~~~~

\newpage

\begin{figure}[hbtp]
\begin{center}
\includegraphics[width=14cm,angle=180]{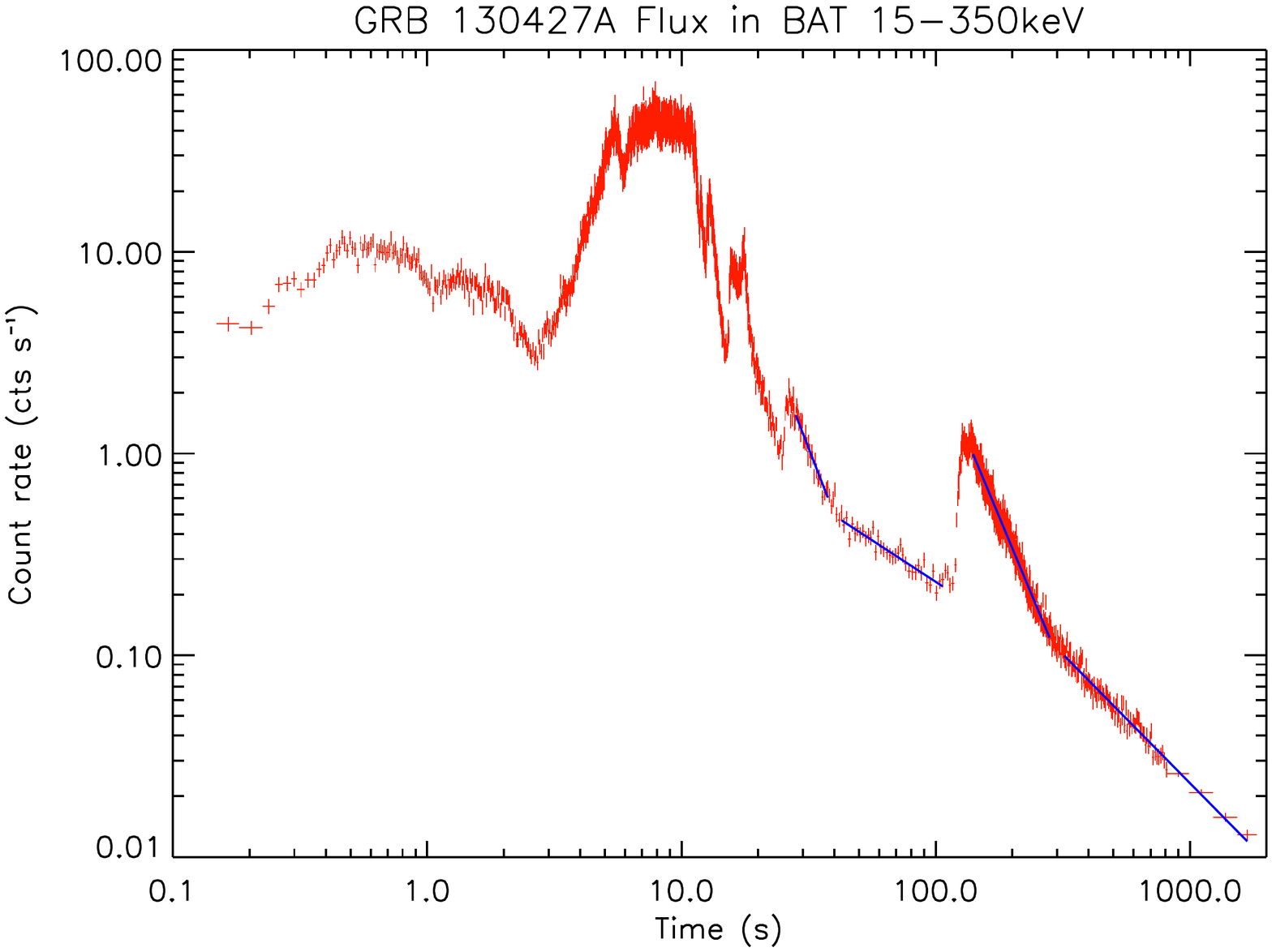}
\end{center}
\end{figure}
\noindent{\bf Figure~S1.} BAT mask-weighted light curve
showing the count rate in the 15--350~keV energy range. The rate has
been corrected for the significant deadtime over the main peak of the
event, with a maximum correction factor of 1.72. The 4 blue curves
shown are power-law fits with indices $\alpha_1$ = 3.25~$\pm$~0.04,
$\alpha_2$ = 0.82~$\pm$~0.04, $\alpha_3$ = 3.02~$\pm$~0.03, and
$\alpha_4$ = 1.28~$\pm$~0.04, from earlier to later times.

\newpage 

\begin{figure}[hbtp]
\begin{center}
\includegraphics[width=16cm]{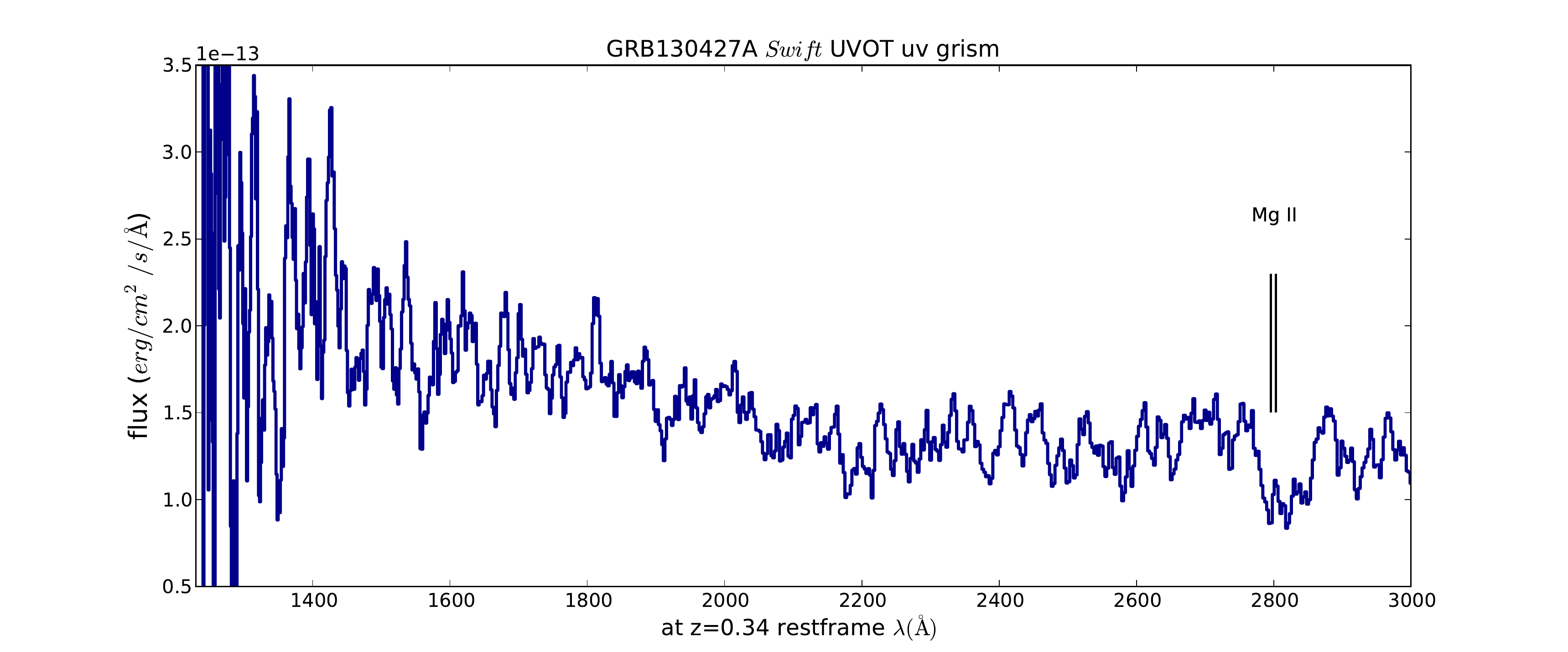}
\end{center}
\end{figure}
\noindent{\bf Figure~S2.} The \sw\ UVOT spectrum with
wavelengths converted to the rest frame of the GRB host. The strongest
absorption feature in the spectrum corresponds to the redshifted Mg~II
2800~\AA\ resonance line. The flux above 2200~\AA\ (in
the rest frame) is overestimated due to second order contamination.

\newpage 

\begin{figure}[hbtp]
\begin{center}
\includegraphics[width=15cm]{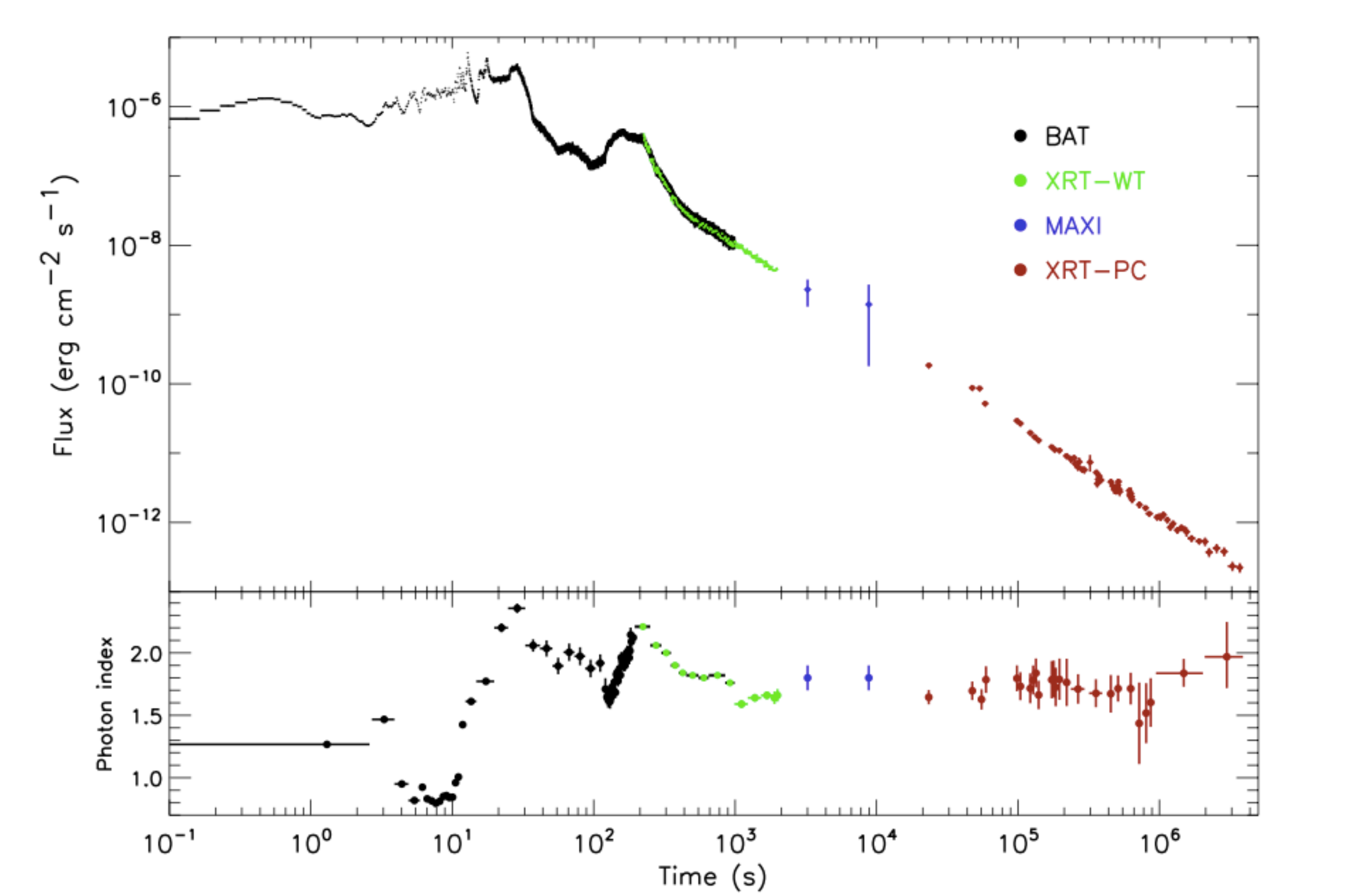}
\end{center}
\end{figure}
\noindent{\footnotesize{\bf Figure~S3.} Top panel:
  composite BAT/XRT light curve starting since 0.3~s after the GBM
  trigger. The XRT light curve in the 0.3$-$10 keV energy range is
  shown together with the BAT light curve extrapolated to the same
  energy range assuming a power-law model in the 15--150 keV energy
  range. The conversion from count rates to fluxes has been performed
  using the spectral parameters listed in Table~S4. Blue
  points correspond to the flux by MAXI obtained assuming $\Gamma$ =
  1.8$\pm$0.1 (see bottom panel) to compute the conversion factor from
  the count rate. Bottom panel: the values of the photon index
  obtained adopting a simple power-law model in the spectral
  analysis.}

\newpage 

\begin{figure}[hbtp]
\begin{center}
\includegraphics[width=14cm]{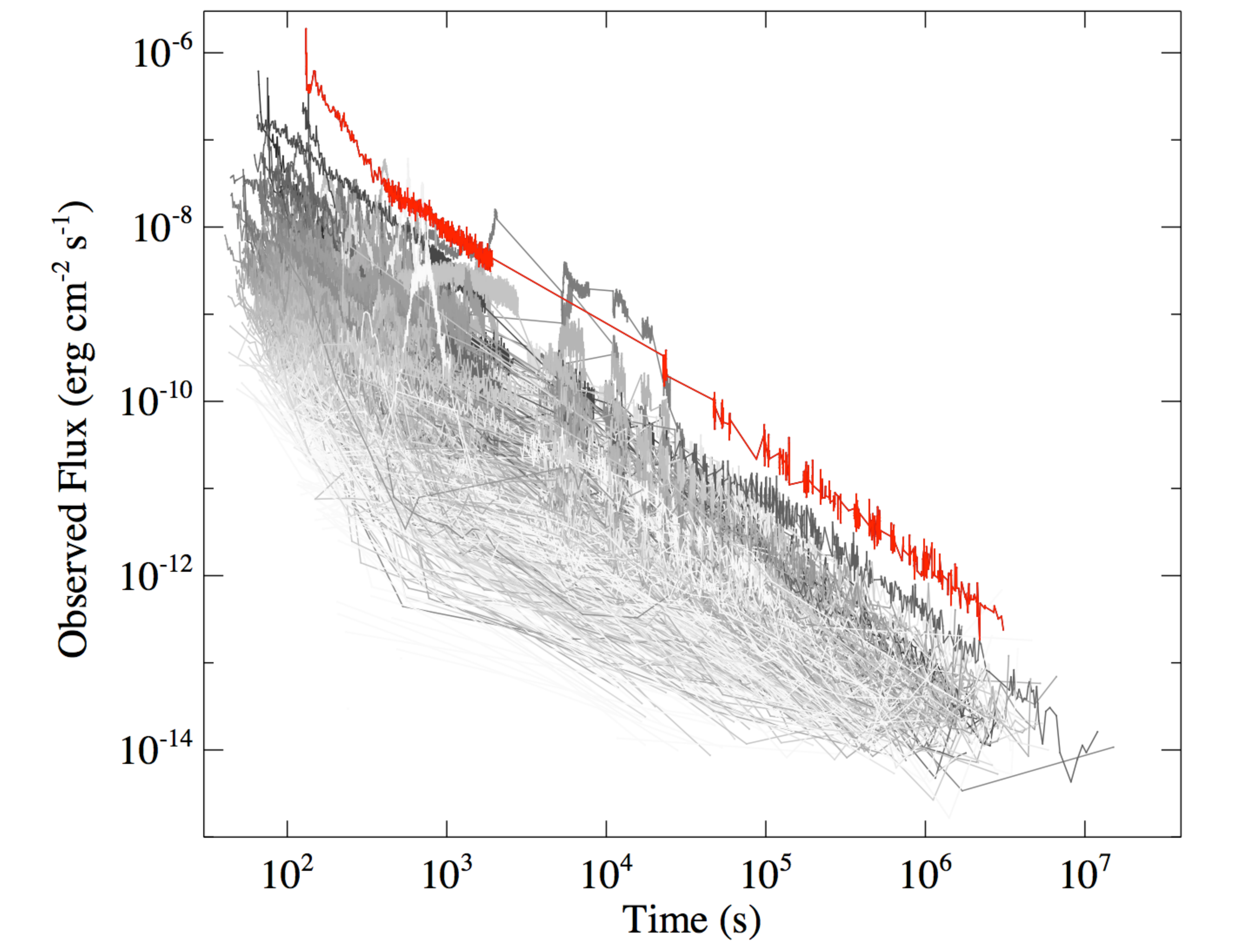}
\end{center}
\end{figure}
\noindent{\footnotesize{\bf Figure~S4.} Comparison of all
  the 627~GRBs observed by XRT (as of June~4, 2013). In terms of the
  observed flux, the X--ray afterglow of \grb, in red color in
  this figure, is the brightest so far observed.}

\newpage 

\begin{figure}[hbtp]
\begin{center}
\includegraphics[width=16cm]{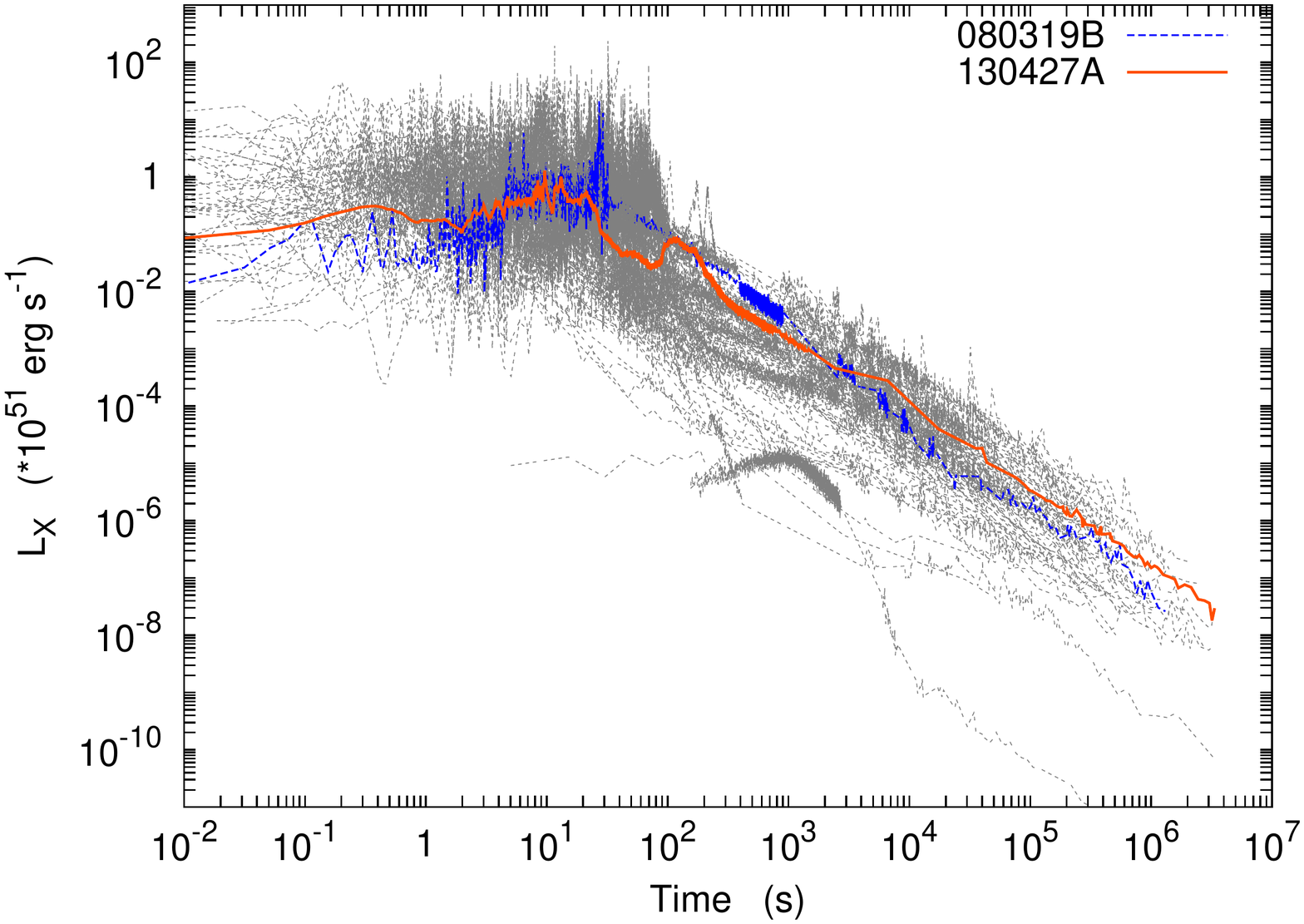}
\end{center}
\end{figure}
\noindent{\footnotesize{\bf Figure~S5.} Rest--frame isotropic X--ray
  luminosity light curves for a selected sample of long, relatively
  bright GRBs (grey curves). For comparison we show in orange and blue
  the rest--frame luminosity for \grb\ and GRB\,080319B ({\it
    naked-eye}), respectively. For each event we also plot the BAT
  $\gamma$-ray luminosity. The behavior and luminosity of \grb\ is
  within the range of long GRBs population at large.}

\newpage 

\begin{figure}[hbtp]
\begin{center}
\includegraphics[width=16cm]{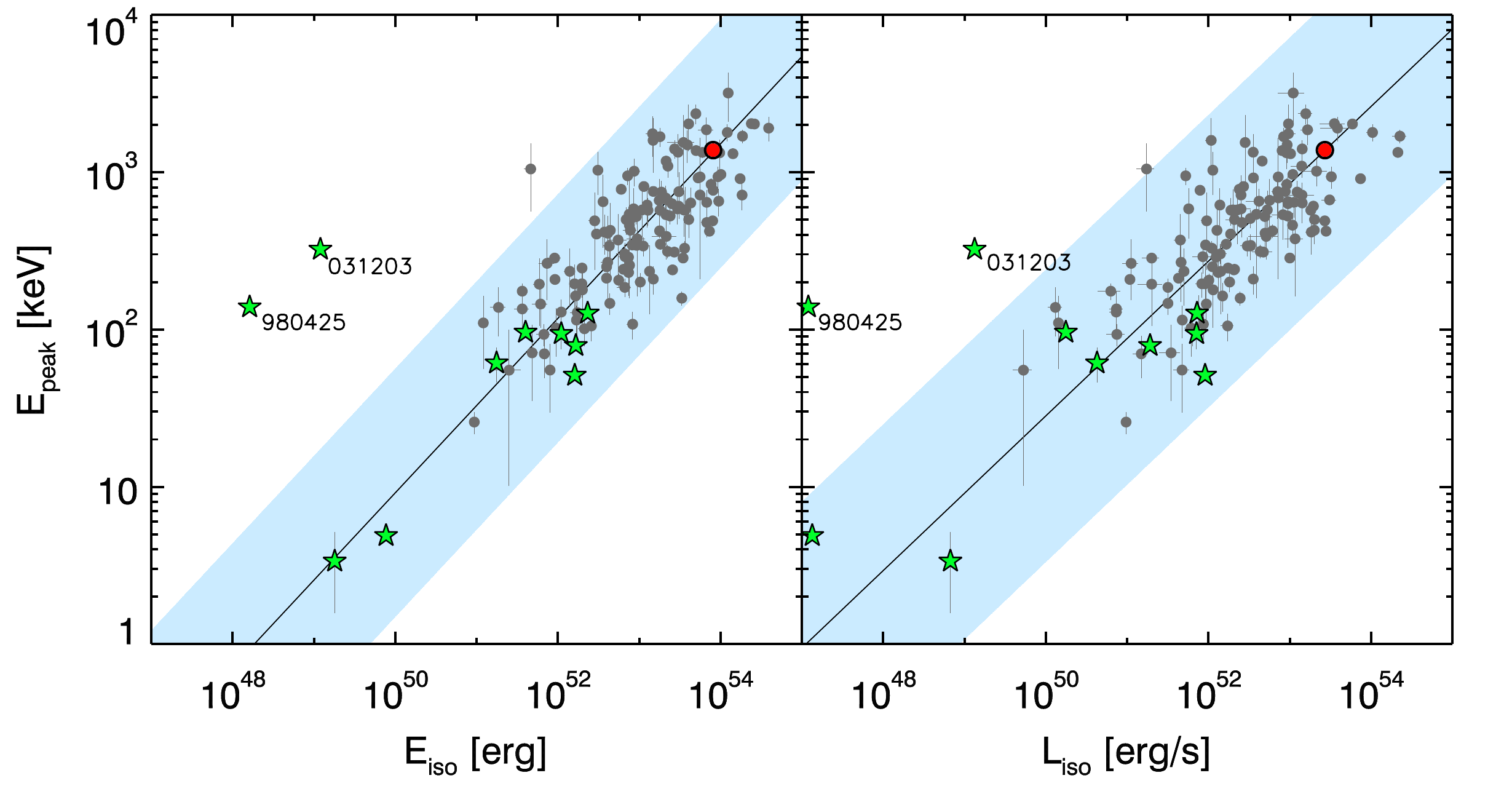}
\end{center}
\end{figure}
\noindent{\footnotesize{\bf Figure~S6.} The Amati (left panel) and
  Yonetoku (right panel) correlations. Grey filled circles refer to
  the sample studied in ({\it 60}). Their power--law
  fit is shown as a solid dark line. Green stars correspond to GRBs
  associated with a spectroscopically confirmed supernova. The shaded
  region represents the 3$\sigma$ scatter of the distribution of
  points around this best fit line. \grb\ (red circle) lies exactly on
  the best fit of both correlations.}

\newpage 

\begin{table}
\footnotesize
\centering
\begin{tabular}{llllllll lllll}
\hline
\hline
Line   &$E_{\rm k}$ & $\epsilon_{\rm e}$ &$\epsilon_{\rm B}$ &$n$  &$p$ 
&$\theta_{\rm j}$  &$t_{\rm k}$ &$\alpha_{\rm k}$ &$t_{\rm e}$ &$\alpha_{\rm e}$
&$t_{\rm B}$ &$\alpha_{\rm B}$  \\
~      & erg/s &           &        & cm$^{-3}$ &   & deg & days &   & days &   & days &   \\
\hline    
dotted &  5e54 &  0.027 & 1.e--5 & 1 & 2.3 & $3.4^\circ$ & ... &  ...  & ... & ... & ... & ... \\
solid  &  5e54 &   0.027 & 1.e--5 & 1 & 2.3 & $3.4^\circ$ & 0.2 & --0.8 & 0.8 & 0.6 & 0.8 & 0.5 \\
\hline
\hline
\end{tabular}
\vskip 0.4 true cm
\noindent{\footnotesize{\bf Table~S10.} List of parameters
  used for the model plotted in the Figure~S7. We assumed
  $k = (t/t_{\rm k})^{\alpha_{\rm k}}$ after $t_{\rm k}$ (and same
  functions for $\epsilon_{\rm e} $ and $\epsilon_{\rm B})$.}
\end{table}

\begin{figure}
\hskip -1 cm
\includegraphics[height=0.5\textheight]{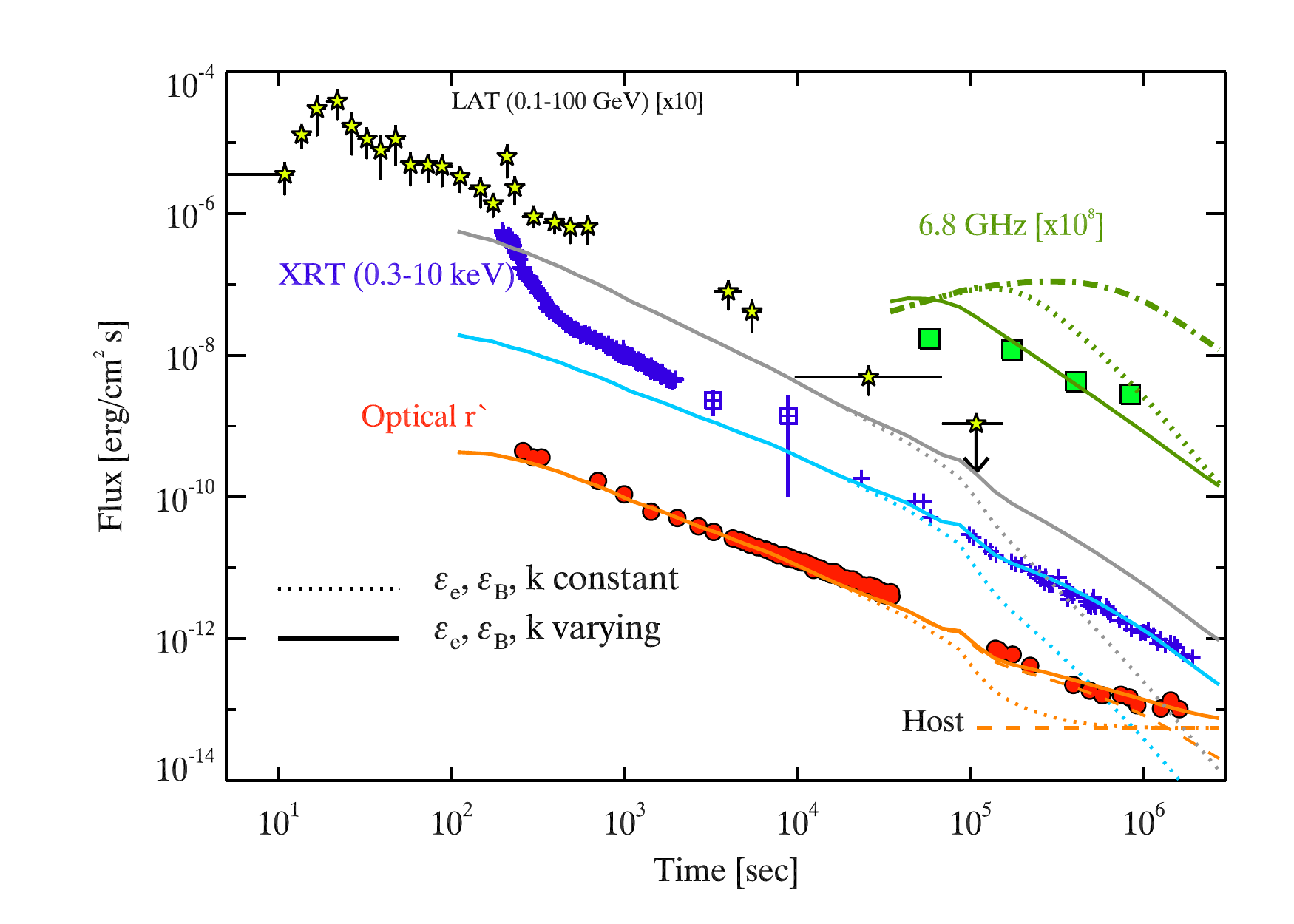}
\vskip -0.3 cm
\end{figure}
\noindent{\footnotesize{\bf Figure~S7.} The optical, X--ray,
  $\gamma$--ray and radio light curves are interpreted as forward
  shock afterglow synchrotron emission, as derived applying the van
  Eerten et al. model. Dotted lines corresponds to the first line in
  Table~S10, namely $\epsilon_{\rm e}$, $\epsilon_{\rm B}$ and the
  fraction $k$ of accelerated electrons are constant. The solid line
  corresponds to the model when varying these microphysical
  parameters.  The dot--dashed line shown only for the radio band,
  corresponds to the model with only $\epsilon_{\rm e}$ and
  $\epsilon_{\rm B}$ varying in time and with constant $k$.}

\newpage 

\begin{table}
\begin{center}
\footnotesize
\begin{tabular}{lllrrr}
\hline
Date       &  ObsID      & Mode &   \tsta   &   \tsto  & Exposure \\ 
\hline 
\hline
2013-04-27 & 00554620000 &  WT  &     195.0 &   2021.0 &  1826.0  \\ 
2013-04-27 & 00554620001 &  PC  &   22895.7 &  23992.1 &  1091.3  \\ 
2013-04-27 & 00554620002 &  PC  &   46969.2 &  48157.6 &  1183.7  \\ 
2013-04-27 & 00554620003 &  PC  &   53000.0 &  57444.6 &  1166.2  \\ 
2013-04-28 & 00554620010 &  PC  &   58984.1 &  59841.1 &   616.8  \\ 
2013-04-28 & 00554620011 &  PC  &   86273.6 &  86308.9 &    34.9  \\ 
2013-04-28 & 00554620012 &  PC  &   97700.2 &  98876.9 &  1171.2  \\ 
2013-04-28 & 00554620013 &  PC  &  103473.2 & 104649.1 &  1171.2  \\ 
2013-04-28 & 00554620014 &  PC  &  121583.6 & 122777.0 &  1188.7  \\ 
2013-04-28 & 00554620015 &  PC  &  127497.0 & 128688.0 &  1186.2  \\ 
2013-04-28 & 00554620016 &  PC  &  133486.8 & 134678.1 &  1186.2  \\ 
2013-04-28 & 00554620017 &  PC  &  139453.2 & 140626.6 &  1168.7  \\ 
2013-04-29 & 00554620018 &  PC  &  172691.4 & 173887.4 &  1191.2  \\ 
2013-04-29 & 00554620019 &  PC  &  178457.6 & 179653.6 &  1191.2  \\ 
2013-04-29 & 00554620020 &  PC  &  184342.2 & 185538.2 &  1191.2  \\ 
2013-04-29 & 00554620021 &  PC  &  195806.2 & 197002.1 &  1191.2  \\ 
2013-04-29 & 00554620022 &  PC  &  214255.0 & 214282.9 &    27.4  \\ 
2013-04-29 & 00554620023 &  PC  &  220516.4 & 226200.7 &  1163.7  \\ 
2013-04-30 & 00554620024 &  PC  &  236492.8 & 284417.9 &  2235.0  \\ 
2013-04-30 & 00554620025 &  PC  &  231962.4 & 295404.9 &  1263.6  \\ 
2013-05-01 & 00554620026 &  PC  &  318431.4 & 387978.9 &  2949.3  \\ 
2013-05-01 & 00554620027 &  PC  &  323916.2 & 393026.9 &  2861.8  \\ 
2013-05-02 & 00554620028 &  PC  &  443986.8 & 480197.1 &  2474.8  \\ 
2013-05-02 & 00554620029 &  PC  &  450022.2 & 462502.1 &  1033.8  \\ 
2013-05-03 & 00554620030 &  PC  &  495862.2 & 520641.9 &  4879.1  \\ 
2013-05-03 & 00554620031 &  PC  &  501622.2 & 525617.9 &  4033.1  \\ 
2013-05-04 & 00554620032 &  PC  &  611311.6 & 636145.9 &  4869.6  \\ 
2013-05-04 & 00554620033 &  PC  &  617072.2 & 642158.1 &  3628.5  \\ 
2013-05-05 & 00554620034 &  PC  &  711244.4 & 734065.1 &   948.9  \\ 
2013-05-05 & 00554620035 &  PC  &  717012.8 & 728662.1 &   946.4  \\ 
\hline
\end{tabular}
\end{center}
\end{table}
\noindent{\footnotesize{\bf Table~S1.} List of XRT Observations.}

\newpage 

\begin{table}
\begin{center}
\footnotesize
\begin{tabular}{lllrrr}
\hline
Date       &  ObsID      & Mode &   \tsta   & \tsto      & Exposure \\ 
\hline 
\hline
2013-05-06 & 00554620036 &  PC  &  790108.6 &   815075.0 &  2577.2  \\ 
2013-05-06 & 00554620037 &  PC  &  795861.2 &   820847.0 &  1872.9  \\ 
2013-05-07 & 00554620038 &  PC  &  842938.6 &   890245.4 &  3316.4  \\ 
2013-05-07 & 00554620039 &  PC  &  848698.4 &   896015.6 &  3875.7  \\ 
2013-05-08 & 00554620040 &  PC  &  946138.4 &  1005185.9 &  3338.8  \\ 
2013-05-08 & 00554620041 &  PC  &  963096.0 &   999986.7 &  4597.4  \\ 
2013-05-09 & 00554620042 &  PC  & 1015070.2 &  1044940.6 &  5096.9  \\ 
2013-05-09 & 00554620043 &  PC  & 1009222.2 &  1034321.2 &  4382.7  \\ 
2013-05-10 & 00554620044 &  PC  & 1101436.4 &  1137558.1 &  3211.5  \\ 
2013-05-10 & 00554620045 &  PC  & 1130278.0 &  1170759.0 &  3109.1  \\ 
2013-05-11 & 00554620046 &  PC  & 1199567.8 &  1223006.9 &  3668.5  \\ 
2013-05-11 & 00554620047 &  PC  & 1205387.8 &  1230084.8 &  4922.1  \\ 
2013-05-12 & 00554620048 &  PC  & 1268802.2 &  1291864.2 &  3238.9  \\ 
2013-05-12 & 00554620049 &  PC  & 1274530.4 &  1297614.9 &  3081.6  \\ 
2013-05-13 & 00554620050 &  PC  & 1356452.8 &  1398086.9 &  2839.4  \\ 
2013-05-13 & 00554620051 &  PC  & 1362212.8 &  1403733.9 &  2205.1  \\ 
2013-05-14 & 00554620052 &  PC  & 1441798.4 &  1472380.9 &  5084.4  \\ 
2013-05-14 & 00554620053 &  PC  & 1447558.0 &  1477901.1 &  4430.1  \\ 
2013-05-15 & 00554620054 &  PC  & 1535337.9 &  1580847.9 &  3793.3  \\ 
2013-05-15 & 00554620055 &  PC  & 1529578.2 &  1575387.9 &  2739.5  \\ 
2013-05-16 & 00554620056 &  PC  & 1614928.2 &  1689963.7 &  2966.7  \\ 
2013-05-16 & 00554620057 &  PC  & 1620680.4 &  1695761.2 &  2931.8  \\ 
2013-05-17 & 00554620058 &  PC  & 1707466.4 &  1759866.0 &   571.8  \\ 
2013-05-17 & 00554620059 &  PC  & 1701445.8 &  1713010.1 &    77.4  \\ 
2013-05-18 & 00554620060 &  PC  & 1789327.4 &  1847062.2 &  3149.0  \\ 
2013-05-18 & 00554620061 &  PC  & 1793698.4 &  1852829.5 &  3151.5  \\ 
2013-05-19 & 00554620062 &  PC  & 1880294.8 &  1950499.0 &  3316.3  \\ 
2013-05-19 & 00554620063 &  PC  & 1874537.6 &  1957165.6 &  4520.0  \\ 
2013-05-20 & 00554620064 &  PC  & 2007777.3 &  2030161.2 &  2302.4  \\ 
2013-05-20 & 00554620065 &  PC  & 2002047.2 &  2025770.0 &  2372.4  \\ 
\hline
\end{tabular}
\end{center}
\end{table}
\noindent{\footnotesize{\bf Table~S1.} Continued.}

\newpage 

\begin{table}
\begin{center}
\footnotesize
\begin{tabular}{lllrrr}
\hline
Date       &  ObsID      & Mode &   \tsta   & \tsto      & Exposure \\ 
\hline 
\hline
2013-05-21 & 00554620066 &  PC  & 2093589.6 &  2128184.9 &  4123.0  \\
2013-05-21 & 00554620067 &  PC  & 2099349.3 &  2122434.4 &  1470.9  \\
2013-05-22 & 00554620068 &  PC  & 2180242.6 &  2215409.1 &  1815.5  \\
2013-05-22 & 00554620069 &  PC  & 2185857.8 &  2210360.0 &  4065.6  \\
2013-05-24 & 00554620070 &  PC  & 2358988.2 &  2382118.1 &  1643.2  \\
2013-05-24 & 00554620071 &  PC  & 2364683.4 &  2389163.0 &  4567.5  \\
2013-05-26 & 00554620072 &  PC  & 2520513.0 &  2543605.9 &   112.4  \\
2013-05-26 & 00554620073 &  PC  & 2514753.2 &  2537833.9 &   127.4  \\
2013-05-28 & 00554620074 &  PC  & 2687649.2 &  2712052.5 &  2584.7  \\
2013-05-28 & 00554620075 &  PC  & 2693409.2 &  2723792.9 &  3341.4  \\
2013-05-30 & 00554620076 &  PC  & 2838150.4 &  2873300.8 &  2517.3  \\
2013-05-30 & 00554620077 &  PC  & 2843910.2 &  2878100.5 &   182.3  \\
2013-05-31 & 00554620078 &  PC  & 2958738.2 &  2970872.2 &  1530.9  \\
2013-06-01 & 00554620079 &  PC  & 3034154.0 &  3058480.6 &  4260.4  \\
2013-06-01 & 00554620080 &  PC  & 3039807.5 &  3064297.5 &  3498.7  \\ 
2013-06-03 & 00554620081 &  PC  & 3207534.7 &  3219986.9 &   964.0  \\
2013-06-03 & 00554620082 &  PC  & 3196010.0 &  3224848.9 &   112.4  \\
2013-06-05 & 00554620083 &  PC  & 3385650.6 &  3426578.8 &  2632.1  \\
2013-06-05 & 00554620084 &  PC  & 3380735.1 &  3420857.9 &  3184.0  \\
2013-06-07 & 00554620085 &  PC  & 3524157.8 &  3548402.3 &  3611.1  \\
2013-06-07 & 00554620086 &  PC  & 3518412.8 &  3542624.9 &  3443.8  \\
2013-06-09 & 00554620087 &  PC  & 3691594.0 &  3726982.5 &  1970.4  \\
2013-06-09 & 00554620088 &  PC  & 3697361.2 &  3698268.8 &   904.0  \\
2013-06-11 & 00554620089 &  PC  & 3870322.0 &  3906218.0 &  2627.1  \\
2013-06-11 & 00554620090 &  PC  & 3864659.2 &  3899309.9 &  1223.7  \\ 
\hline
\end{tabular}
\end{center}
\end{table}
\noindent{\footnotesize{\bf Table~S1.} Continued.}

\newpage 

\begin{table}
\footnotesize
\begin{center}
\begin{tabular}{rrrrr}
\hline
Mode  &  \tsta  &  \tsto  &    r  &  R    \\ 
\hline                                                              
   WT  &     195  &      250 & 30 & 60    \\
   WT  &     250  &      300 & 20 & 60    \\   
   WT  &     300  &      350 & 15 & 60    \\ 
   WT  &     350  &      400 & 12 & 60    \\ 
   WT  &     400  &      450 & 10 & 60    \\ 
   WT  &     450  &      550 &  9 & 60    \\ 
   WT  &     550  &      650 &  8 & 60    \\ 
   WT  &     650  &      850 &  7 & 60    \\ 
   WT  &     850  &     1150 &  4 & 60    \\ 
   WT  &    1150  &     1851 &  3 & 60    \\ 
~\\
   PC  &   22874  &    54172 &  4 & 32    \\
   PC  &   57429  &    59534 &  2 & 27    \\
\hline
\end{tabular}
\end{center}
\end{table}
\noindent{\footnotesize{\bf Table~S2.} The value,
  expressed in pixels (1 pixel = 2.36\as), of the inner r and outer R
  radius of the annular extraction region adopted to take into account
  the pile-up corrections, when needed.}

\newpage 

\begin{table*}
\footnotesize
\begin{center}
\begin{tabular}{rrccc}
\hline
\tsta      & Time interval &         Flux           &        Magnitude     & Filter \\
\hline
    s        &      s     &           mJy           &         mag          &        \\
\hline
      432.9  &      19.4  &  55.86   $\pm$   1.78   &  12.10  $\pm$  0.04  &  $v$   \\
      606.5  &      19.7  &  38.62   $\pm$   1.18   &  12.50  $\pm$  0.03  &  $v$   \\
      779.2  &      19.7  &  31.23   $\pm$   0.98   &  12.73  $\pm$  0.04  &  $v$   \\
     1104.7  &      19.7  &  19.01   $\pm$   0.67   &  13.27  $\pm$  0.04  &  $v$   \\
     1278.7  &      19.7  &  16.55   $\pm$   0.60   &  13.42  $\pm$  0.04  &  $v$   \\
     1450.9  &      19.7  &  13.92   $\pm$   0.54   &  13.61  $\pm$  0.04  &  $v$   \\
     1623.2  &      19.2  &  11.90   $\pm$   0.50   &  13.78  $\pm$  0.05  &  $v$   \\
     1795.5  &     176.8  &  10.87   $\pm$   0.42   &  13.88  $\pm$  0.04  &  $v$   \\
     1972.3  &      15.8  &  10.77   $\pm$   0.51   &  13.89  $\pm$  0.05  &  $v$   \\
   221178.8  &      79.8  &  0.042   $\pm$   0.025  &  19.91  $\pm$  0.54  &  $v$   \\
   271227.3  &   24196.7  &  0.019   $\pm$   0.016  &  20.72  $\pm$  0.69  &  $v$   \\
   358892.1  &   23111.9  &  0.050   $\pm$   0.009  &  19.70  $\pm$  0.19  &  $v$   \\
   462312.3  &   46771.8  &  0.020   $\pm$   0.006  &  20.68  $\pm$  0.32  &  $v$   \\
   520269.5  &     394.6  &  0.017   $\pm$   0.010  &  20.84  $\pm$  0.53  &  $v$   \\
   612573.7  &   23590.4  &  0.021   $\pm$   0.006  &  20.66  $\pm$  0.30  &  $v$   \\
   797123.6  &   64580.4  &  0.016   $\pm$   0.006  &  20.94  $\pm$  0.37  &  $v$   \\
   884582.0  &   79902.1  &  0.006   $\pm$   0.009  &  22.03  $\pm$  1.07  &  $v$   \\
   981779.5  &   63164.5  &  0.012   $\pm$   0.004  &  21.23  $\pm$  0.37  &  $v$   \\
  1102735.3  &  110088.8  &  0.015   $\pm$   0.005  &  21.00  $\pm$  0.35  &  $v$   \\
  1270046.2  &  128057.8  &  0.015   $\pm$   0.005  &  20.97  $\pm$  0.35  &  $v$   \\
  1443096.2  &   29287.8  &  0.018   $\pm$   0.006  &  20.81  $\pm$  0.33  &  $v$   \\
  1615695.2  &     208.9  &  0.035   $\pm$   0.015  &  20.11  $\pm$  0.42  &  $v$   \\
  1800561.6  &  149962.6  &  0.014   $\pm$   0.005  &  21.08  $\pm$  0.39  &  $v$   \\
  2008769.3  &  206654.8  &  0.013   $\pm$   0.004  &  21.15  $\pm$  0.34  &  $v$   \\
\hline
\end{tabular}
\end{center}
\end{table*}
\noindent{\footnotesize{\bf Table~S3.} UVOT Observation
  log, reporting the interval of time over which observations have
  been collected since \tsta. Observations where reliable photometry
  could not be extracted are not included in the table. Magnitudes
  have not been corrected for Galactic absorption along the line of
  sight (E$_{B-V}$ = 0.020 mag; ({\it 61})). The
  estimate extinctions in the different filters are: A$_{\rm v}$ =
  0.062~mag, A$_{\rm b}$ = 0.079~mag, A$_{\rm u}$ = 0.096~mag, A$_{\rm
    w1}$ = 0.132~mag, A$_{\rm m2}$ = 0.193~mag, and A$_{\rm w2}$ =
  0.175~mag. Corrected magnitudes have been converted into flux
  densities, F$_{\nu}$, following ({\it 62})}.

\newpage 

\begin{table*}
\footnotesize
\begin{center}
\begin{tabular}{rrccc}
\hline
\tsta        & Time interval &          Flux         &    Magnitude       & Filter \\
\hline
    s        &       s       &          mJy          &       mag          &        \\
\hline
      531.3  &      19.8     & 111.71  $\pm$ 41.54   &  11.48 $\pm$ 0.56  &  $b$   \\
      705.2  &      19.7     &  30.25  $\pm$  1.04   &  12.90 $\pm$ 0.04  &  $b$   \\
      877.1  &      19.7     &  23.12  $\pm$  0.71   &  13.19 $\pm$ 0.04  &  $b$   \\
     1202.9  &      19.7     &  16.05  $\pm$  0.48   &  13.59 $\pm$ 0.03  &  $b$   \\
     1376.9  &      19.7     &  13.11  $\pm$  0.39   &  13.81 $\pm$ 0.03  &  $b$   \\
     1549.3  &      19.7     &  11.28  $\pm$  0.34   &  13.97 $\pm$ 0.04  &  $b$   \\
\hline
\end{tabular}
\end{center}
\end{table*}
\noindent{\footnotesize{\bf Table~S3.} Continued}.

\newpage

\begin{table*}
\footnotesize
\begin{center}
\begin{tabular}{rrccc}
\hline
\tsta      & Time interval &            Flux       &        Magnitude       &   Filter   \\
\hline
    s      &       s       &          mJy          &           mag          &            \\
\hline
      506.3  &      19.8  &  52.75   $\pm$  18.58   &  11.19  $\pm$  0.35  &  $u$   \\
      680.3  &       5.9  &  25.06   $\pm$   1.61   &  12.00  $\pm$  0.07  &  $u$   \\
      852.4  &      19.7  &  21.97   $\pm$   0.73   &  12.14  $\pm$  0.04  &  $u$   \\
     1178.1  &      19.7  &  14.55   $\pm$   0.44   &  12.59  $\pm$  0.04  &  $u$   \\
     1352.2  &      19.7  &  11.68   $\pm$   0.34   &  12.83  $\pm$  0.03  &  $u$   \\
     1524.0  &      19.7  &  10.67   $\pm$   0.31   &  12.93  $\pm$  0.03  &  $u$   \\
\hline
\end{tabular}
\end{center}
\end{table*}
\noindent{\footnotesize{\bf Table~S3.} Continued}.

\newpage 

\begin{table*}
\footnotesize
\begin{center}
\begin{tabular}{rrccc}
\hline
\tsta      & Time interval &              Flux       &      Magnitude      &  Filter \\
\hline
    s      &     s     &            mJy          &         mag         &         \\
\hline
    556.8  &     20.0  &     52.22 $\pm$   13.70 &   11.52 $\pm$  0.27 & $white$ \\
    729.7  &     20.0  &     28.38 $\pm$   11.16 &   12.18 $\pm$  0.39 & $white$ \\
    924.4  &    150.0  &     18.79 $\pm$    4.55 &   12.63 $\pm$  0.26 & $white$ \\
 271152.9  &  24121.4  &     0.041 $\pm$   0.003 &   19.28 $\pm$  0.09 & $white$ \\
 358629.3  &  23141.3  &     0.028 $\pm$   0.002 &   19.67 $\pm$  0.06 & $white$ \\
 462042.5  &  46641.4  &     0.019 $\pm$   0.001 &   20.10 $\pm$  0.07 & $white$ \\
 519834.3  &    428.8  &     0.018 $\pm$   0.002 &   20.13 $\pm$  0.12 & $white$ \\
 612153.5  &  23645.9  &     0.014 $\pm$   0.001 &   20.43 $\pm$  0.09 & $white$ \\
 728641.2  &  68476.1  &     0.013 $\pm$   0.002 &   20.52 $\pm$  0.16 & $white$ \\
 849450.8  &  35125.5  &     0.011 $\pm$   0.002 &   20.68 $\pm$  0.13 & $white$ \\
 963818.3  &  80879.6  &     0.008 $\pm$   0.001 &   21.04 $\pm$  0.09 & $white$ \\
1102303.2  & 110109.7  &     0.008 $\pm$   0.001 &   21.03 $\pm$  0.11 & $white$ \\
1269632.1  & 128173.5  &     0.009 $\pm$   0.001 &   20.88 $\pm$  0.10 & $white$ \\
1442664.2  &  93347.3  &     0.008 $\pm$   0.001 &   21.00 $\pm$  0.12 & $white$ \\
1615440.4  &    248.8  &     0.004 $\pm$   0.002 &   21.88 $\pm$  0.43 & $white$ \\
1800201.6  & 150031.8  &     0.007 $\pm$   0.001 &   21.19 $\pm$  0.12 & $white$ \\
2008439.3  & 206801.2  &     0.006 $\pm$   0.001 &   21.28 $\pm$  0.11 & $white$ \\
\hline
\end{tabular}
\end{center}
\end{table*}
\noindent{\footnotesize{\bf Table~S3.} Continued}.

\newpage 

\begin{table*}
\footnotesize
\begin{center}
\begin{tabular}{rrccc}
\hline
\tsta      & Time interval &            Flux       &        Magnitude       &  Filter     \\
\hline
    s      &    s     &          mJy          &           mag          &            \\
\hline
      353.1  &      50.0  &  34.74   $\pm$   6.30   &  11.16  $\pm$  0.20  &  $w1$  \\
      482.0  &      19.7  &  32.06   $\pm$   1.03   &  11.25  $\pm$  0.04  &  $w1$  \\
      656.0  &      19.7  &  21.16   $\pm$   0.60   &  11.70  $\pm$  0.03  &  $w1$  \\
      828.1  &      19.7  &  16.80   $\pm$   0.47   &  11.95  $\pm$  0.03  &  $w1$  \\
     1153.9  &      19.7  &  10.77   $\pm$   0.32   &  12.43  $\pm$  0.04  &  $w1$  \\
     1327.9  &      19.7  &   9.71   $\pm$   0.29   &  12.54  $\pm$  0.04  &  $w1$  \\
\hline
\end{tabular}
\end{center}
\end{table*}
\noindent{\footnotesize{\bf Table~S3.} Continued}.

\newpage 

\begin{table*}
\scriptsize
\begin{center}
\begin{tabular}{rrccc}
\hline
\tsta      & Time interval &            Flux       &        Magnitude       &  Filter \\
\hline
    s      &    s     &          mJy          &           mag          &       \\
\hline
      353.1  &      50.0  &  34.13   $\pm$   4.96   &  11.04  $\pm$  0.17  &  $m2$  \\
      457.4  &      19.7  &  30.34   $\pm$   0.87   &  11.17  $\pm$  0.04  &  $m2$  \\
      630.8  &      19.7  &  21.01   $\pm$   0.59   &  11.57  $\pm$  0.03  &  $m2$  \\
      803.4  &      19.7  &  16.65   $\pm$   0.48   &  11.82  $\pm$  0.04  &  $m2$  \\
     1129.2  &      19.7  &  10.41   $\pm$   0.34   &  12.33  $\pm$  0.04  &  $m2$  \\
     1303.1  &      19.7  &   9.337  $\pm$   0.31   &  12.45  $\pm$  0.04  &  $m2$  \\
\hline
\end{tabular}
\end{center}
\end{table*}
\noindent{\footnotesize{\bf Table~S3.} Continued}.

\newpage 

\begin{table*}
\scriptsize
\begin{center}
\begin{tabular}{rrccc}
\hline
\tsta      & Time interval &            Flux       &        Magnitude       &   Filter     \\
\hline
    s      &    s     &          mJy          &           mag          &            \\
\hline
      353.1  &      50.0  &  29.93   $\pm$   5.21   &  11.16  $\pm$  0.20  &  $w2$  \\
      408.6  &      19.7  &  29.27   $\pm$   0.91   &  11.18  $\pm$  0.04  &  $w2$  \\
      582.1  &      19.7  &  19.99   $\pm$   0.55   &  11.60  $\pm$  0.03  &  $w2$  \\
      754.9  &      19.7  &  13.85   $\pm$   0.37   &  12.00  $\pm$  0.03  &  $w2$  \\
     1080.3  &      19.7  &  10.43   $\pm$   0.29   &  12.30  $\pm$  0.04  &  $w2$  \\
     1254.4  &      19.7  &   7.98   $\pm$   0.23   &  12.60  $\pm$  0.04  &  $w2$  \\
\hline
\end{tabular}
\end{center}
\end{table*}
\noindent{\footnotesize{\bf Table~S3.} Continued}.

\newpage 

\begin{table*}
\begin{center}
\footnotesize
\begin{tabular}{crrcccc}
\hline
 Instrument  &   \tsta    &    \tsto  &     N$_{\rm H}$     &    $\Gamma$      &        Normalization                    & $\chi^2$ (d.o.f.) \\ 
\hline                                                                                    
             &      s     &     s     & 10$^{22}$ cm$^{-2}$ &                  & ph cm$^{-2}$ s$^{-1}$ keV$^{-1}$ @ 50~keV  &             \\ 
\hline                                                                                    
BAT          &     0.0    &      2.8  &        -          &  1.27 $\pm$ 0.03  &           0.45  $\pm$ 0.01              &  0.98 (57)  \\ 
BAT          &     2.8    &      4.0  &        -          &  1.47 $\pm$ 0.04  &           0.39  $\pm$ 0.01              &  1.74 (57)  \\  
BAT          &     4.0    &      5.0  &        -          &  0.95 $\pm$ 0.03  &           1.15  $\pm$ 0.02              &  1.10 (57)  \\ 
BAT          &     5.0    &      6.0  &        -          &  0.82 $\pm$ 0.02  &           2.36  $\pm$ 0.04              &  1.91 (57)  \\ 
BAT          &     6.0    &      6.5  &        -          &  0.92 $\pm$ 0.03  &           2.22  $\pm$ 0.04              &  1.48 (57)  \\ 
BAT          &     6.5    &      7.0  &        -          &  0.83 $\pm$ 0.03  &           2.87  $\pm$ 0.05              &  1.49 (57)  \\
\hline
\end{tabular}
\end{center}
\end{table*}
\noindent{\footnotesize{\bf Table~S4.} Parameters of the spectral
  analysis of \sw\ data. The values of a simple power-law fit are
  reported for different time intervals. BAT and XRT spectra were
  simultaneously fitted for 195~s $<$ t $<$ 990~s.}

 \newpage 

\begin{table*}
\begin{center}
\footnotesize
\begin{tabular}{crrcccc}
\hline
 Instrument  &   \tsta    &    \tsto  &     N$_{\rm H}$     &    $\Gamma$      &        Normalization                    &    $\chi^2$ (d.o.f.)    \\ 
\hline                                                  
             &      s     &     s     & 10$^{22}$ cm$^{-2}$ &                  & ph cm$^{-2}$ s$^{-1}$ keV$^{-1}$ @ 50~keV &                          \\ 
\hline                                                  
BAT          &   127.7    &    129.7  &         -         &  1.67 $\pm$ 0.10  &           0.072 $\pm$ 0.004            &               0.72 (57)  \\
BAT          &   129.7    &    131.7  &         -         &  1.61 $\pm$ 0.09  &           0.074 $\pm$ 0.004            &               1.04 (57)  \\
BAT          &   131.7    &    133.7  &         -         &  1.64 $\pm$ 0.10  &           0.065 $\pm$ 0.004            &               0.72 (57)  \\
BAT          &   133.7    &    135.7  &         -         &  1.73 $\pm$ 0.09  &           0.066 $\pm$ 0.003            &               1.25 (57)  \\
BAT          &   135.7    &    137.7  &         -         &  1.69 $\pm$ 0.08  &           0.072 $\pm$ 0.003            &               0.88 (57)  \\
BAT          &   137.7    &    139.7  &         -         &  1.69 $\pm$ 0.08  &           0.072 $\pm$ 0.003            &               1.14 (57)  \\
\hline
\end{tabular}
\end{center}
\end{table*}
\noindent{\footnotesize{\bf Table~S4.} Continued.}

\newpage 

\begin{table*}
\begin{center}
\footnotesize
\begin{tabular}{crrccccc}
\hline
 Instrument  &   \tsta    &    \tsto  &     N$_{\rm H}$     &    $\Gamma$      &        Normalization                  &    $\chi^2$ (d.o.f.)    \\ 
\hline                      
             &      s     &     s     & 10$^{22}$ cm$^{-2}$ &                  & ph cm$^{-2}$ s$^{-1}$ keV$^{-1}$ @ 1~keV &                          \\ 
\hline                      
~\\                                                                                               
BAT + XRT-WT &   195      &    250    &  0.30 $\pm$ 0.02  &  2.21 $\pm$ 0.01  &         79    $\pm$ 2                  &  1.59 (331)  \\
BAT + XRT-WT &   250      &    300    &  0.13 $\pm$ 0.02  &  2.06 $\pm$ 0.02  &         23    $\pm$ 1                  &  1.12 (299)  \\
BAT + XRT-WT &   300      &    350    &  0.06 $\pm$ 0.02  &  2.00 $\pm$ 0.02  &         13.7  $\pm$ 0.7                &  1.26 (286)  \\
BAT + XRT-WT &   350      &    400    &  0.04 $\pm$ 0.06  &  1.90 $\pm$ 0.02  &          7.9  $\pm$ 0.4                &  1.20 (247)  \\
BAT + XRT-WT &   400      &    450    &  0.04 $\pm$ 0.02  &  1.84 $\pm$ 0.02  &          5.3  $\pm$ 0.3                &  1.08 (251)  \\
\hline
\end{tabular}
\end{center}
\end{table*}
\noindent{\footnotesize{\bf Table~S4.} Continued.}

\newpage 

\begin{table*}
\scriptsize
\begin{center}
\begin{tabular}{rrccc}
\hline
\tsta      & Exposure &            Flux       &        Magnitude       &  Filter     \\
\hline
    s      &    s     &           mJy         &           mag          &             \\
\hline
    591.1   &       10.0    &  49.12   $\pm$  0.23   &  12.21  $\pm$  0.01    &  $i'$  \\
    838.9   &       30.0    &  37.26   $\pm$  0.17   &  12.51  $\pm$  0.01    &  $i'$  \\
   1165.9   &       60.0    &  22.87   $\pm$  0.10   &  13.04  $\pm$  0.01    &  $i'$  \\
   2317.9   &       180.0   &  10.55   $\pm$  0.05   &  13.88  $\pm$  0.01    &  $i'$  \\
   2914.3   &       120.0   &   8.53   $\pm$  0.04   &  14.11  $\pm$  0.01    &  $i'$  \\
   3595.3   &       180.0   &   6.97   $\pm$  0.03   &  14.33  $\pm$  0.01    &  $i'$  \\
\hline
\end{tabular}
\end{center}
\end{table*}
\noindent{\footnotesize{\bf Table~S5.} Faulkes Telescope
  North observation log. Magnitudes have not been corrected for
  Galactic absorption along the line of sight (E$_{B-V}$ = 0.020~mag;
  ({\it 61})). The estimate extinctions in the
  different filters are: A$_{\rm i'}$ = 0.040~mag, A$_{\rm r'}$
  =~0.052 mag, A$_{\rm V}$ = 0.062~mag, and A$_{\rm B}$ =
  0.079~mag. Corrected magnitudes have been converted into flux
  densities, F$_{\nu}$, following ({\it 62})}.

\newpage 

\begin{table*}
\scriptsize
\begin{center}
\begin{tabular}{rrccc}
\hline
\tsta      & Exposure &            Flux       &        Magnitude       &  Filter     \\
\hline
    s      &    s     &           mJy         &           mag          &             \\
\hline
    310.9   &       10.0    &  91.29  $\pm$  0.17   &  11.553  $\pm$  0.002  &   $r'$  \\
    348.1   &       10.0    &  74.34  $\pm$  0.14   &  11.776  $\pm$  0.002  &   $r'$  \\
    382.9   &       10.0    &  74.27  $\pm$  0.14   &  11.777  $\pm$  0.002  &   $r'$  \\
    754.9   &       30.0    &  34.68  $\pm$  0.06   &  12.604  $\pm$  0.002  &   $r'$  \\
   1047.1   &       60.0    &  22.43  $\pm$  0.04   &  13.077  $\pm$  0.002  &   $r'$  \\
   1477.9   &       120.0   &  12.73  $\pm$  0.02   &  13.692  $\pm$  0.002  &   $r'$  \\
\hline
\end{tabular}
\end{center}
\end{table*}
\noindent{\footnotesize{\bf Table~S5.} Continued.}

\newpage 

\begin{table*}
\scriptsize
\begin{center}
\begin{tabular}{rrccc}
\hline
\tsta      & Exposure &            Flux       &        Magnitude       &  Filter     \\
\hline
    s      &    s     &           mJy         &           mag          &             \\
\hline
    526.3   &  10.0  &  38.22  $\pm$  0.10  &  12.513  $\pm$  0.003  &           $V$  \\
   4462.9   &  10.0  &   4.50  $\pm$  0.06  &  14.836  $\pm$  0.013  &           $V$  \\
   7966.3   &  10.0  &   2.39  $\pm$  0.05  &  15.522  $\pm$  0.021  &           $V$  \\
  11740.3   &  10.0  &   1.63  $\pm$  0.04  &  15.941  $\pm$  0.027  &           $V$  \\
  15568.9   &  10.0  &   1.35  $\pm$  0.05  &  16.145  $\pm$  0.042  &           $V$  \\
~\\
\hline
\end{tabular}
\end{center}
\end{table*}
\noindent{\footnotesize{\bf Table~S5.} Continued.}

\newpage 

\begin{table*}
\footnotesize
\begin{center}
\begin{tabular}{rrccc}
\hline
\tsta      & Exposure &         Flux          &        Magnitude       &  Filter     \\
\hline
    s      &     s    &          mJy          &           mag          &            \\
\hline
 146703.7  &  600.0   &  0.171  $\pm$  0.002  &  18.356  $\pm$  0.010  &  $i'$   \\
 220285.9  &  600.0   &  0.118  $\pm$  0.002  &  18.759  $\pm$  0.012  &  $i'$   \\
 393591.1  &  600.0   &  0.055  $\pm$  0.002  &  19.594  $\pm$  0.027  &  $i'$   \\
 489248.5  &  600.0   &  0.051  $\pm$  0.001  &  19.664  $\pm$  0.023  &  $i'$   \\
 578223.1  &  900.0   &  0.038  $\pm$  0.001  &  19.988  $\pm$  0.034  &  $i'$   \\
 739861.3  &  900.0   &  0.032  $\pm$  0.001  &  20.175  $\pm$  0.036  &  $i'$   \\
\hline
\end{tabular}
\end{center}
\end{table*}
\noindent{\footnotesize{\bf Table~S6.} Liverpool telescope observation
  log. Magnitudes have not been corrected for Galactic absorption
  along the line of sight (E$_{B-V}$ = 0.020~mag; ({\it 61})). The
  estimate extinctions in the different filters are: A$_{\rm i'}$ =
  0.040~mag, and A$_{\rm r'}$ =~0.052 mag. Corrected magnitudes have
  been converted into flux densities, F$_{\nu}$, following ({\it
    62})}.

\newpage

\begin{table*}
\tiny
\begin{center}
\begin{tabular}{rrccc}
\hline
\tsta      & Exposure &         Flux          &        Magnitude       &  Filter    \\
\hline
    s      &    s     &          mJy          &          mag           &            \\
\hline                                           
   8802.7  &   420.0  &  3.40   $\pm$  0.13   &  15.108  $\pm$  0.041  &  $i'$  \\
   9330.7  &   600.0  &  3.03   $\pm$  0.06   &  15.233  $\pm$  0.022  &  $i'$  \\
   9988.9  &   600.0  &  2.88   $\pm$  0.04   &  15.289  $\pm$  0.016  &  $i'$  \\
  10978.9  &  1200.0  &  2.77   $\pm$  0.04   &  15.333  $\pm$  0.015  &  $i'$  \\
  12297.1  &  1200.0  &  2.45   $\pm$  0.04   &  15.464  $\pm$  0.016  &  $i'$  \\
  13614.1  &  1200.0  &  2.14   $\pm$  0.03   &  15.613  $\pm$  0.017  &  $i'$  \\
\hline                                           
\end{tabular}
\end{center}
\end{table*}
\noindent{\scriptsize{\bf Table~S7.} Akeno Observatory observation
  log. Magnitudes have not been corrected for Galactic absorption
  along the line of sight (E$_{B-V}$ = 0.020~mag; ({\it 61})). The
  estimate extinctions in the different filters are: A$_{\rm i'}$ =
  0.040~mag, A$_{\rm r'}$ =~0.052 mag, A$_{\rm g'}$ =
  0.072~mag. Corrected magnitudes have been converted into flux
  densities, F$_{\nu}$, following ({\it 62})}.

\newpage 

\begin{table*}
\footnotesize
\begin{center}
\begin{tabular}{rrccc}
\hline
\tsta      & Exposure &            Flux       &        Magnitude       &  Filter \\
\hline
    s      &    s     &          mJy          &           mag          &         \\
\hline 
   12416.5  &  1800.0  &  2.32   $\pm$  0.05   &  15.525  $\pm$  0.023  &  $i'$  \\
   14475.1  &  1800.0  &  2.22   $\pm$  0.05   &  15.570  $\pm$  0.022  &  $i'$  \\
   16525.3  &  1800.0  &  1.94   $\pm$  0.05   &  15.716  $\pm$  0.028  &  $i'$  \\
   18576.1  &  1800.0  &  1.90   $\pm$  0.08   &  15.741  $\pm$  0.045  &  $i'$  \\
   20770.3  &  2040.0  &  1.70   $\pm$  0.06   &  15.860  $\pm$  0.042  &  $i'$  \\
   25421.5  &  2760.0  &  1.26   $\pm$  0.04   &  16.188  $\pm$  0.037  &  $i'$  \\
\hline
\end{tabular}
\end{center}
\end{table*}
\noindent{\footnotesize{\bf Table~S8.} Okayama Observatory observation
  log. Magnitudes have not been corrected for Galactic absorption
  along the line of sight (E$_{B-V}$ = 0.020~mag; ({\it 61})). The
  estimate extinctions in the different filters are: A$_{\rm i'}$ =
  0.040~mag, A$_{\rm r'}$ =~0.052 mag, and A$_{\rm g'}$ =
  0.072~mag. Corrected magnitudes have been converted into flux
  densities, F$_{\nu}$, following ({\it 62})}.

\newpage 

\begin{table*}
\footnotesize
\begin{center}
\begin{tabular}{rrccc}
\hline
\tsta      & Exposure &            Flux       &        Magnitude       &  Filter \\
\hline
    s      &    s     &          mJy          &           mag          &         \\
\hline
  110172.1  &  1200.0  &  0.225  $\pm$  0.012  &  18.057  $\pm$  0.059  &  $i'$  \\
  112361.5  &  1200.0  &  0.220  $\pm$  0.030  &  18.081  $\pm$  0.150  &  $i'$  \\
  116807.5  &  1500.0  &  0.193  $\pm$  0.017  &  18.225  $\pm$  0.095  &  $i'$  \\
  200004.7  &  1800.0  &  0.099  $\pm$  0.007  &  18.942  $\pm$  0.079  &  $i'$  \\
  202120.3  &  1800.0  &  0.103  $\pm$  0.008  &  18.906  $\pm$  0.083  &  $i'$  \\
  205012.9  &  3180.0  &  0.103  $\pm$  0.008  &  18.902  $\pm$  0.080  &  $i'$  \\
\hline
\end{tabular}
\end{center}
\end{table*}
\noindent{\footnotesize{\bf Table~S9.} Ishigakijima Observatory
  observation log. Magnitudes have not been corrected for Galactic
  absorption along the line of sight (E$_{B-V}$ = 0.020~mag; ({\it
    61})). The estimate extinctions in the different filters are:
  A$_{\rm i'}$ = 0.040~mag, A$_{\rm r'}$ =~0.052 mag, and A$_{\rm g'}$
  = 0.072~mag. Corrected magnitudes have been converted into flux
  densities, F$_{\nu}$, following ({\it 62})}.

\newpage

%

\end{document}